%% file: antiproton.tex
\documentclass[preprint,showpacs,amsmath,amssymb,aps,floats,floatfix,nofootinbib]{revtex4-1}
\usepackage{amsmath,amsfonts,amssymb}
\usepackage{graphicx}
\usepackage{color}
\usepackage[dvipsnames,svgnames,x11names]{xcolor}
\usepackage{slashed} 
\usepackage{url}
\usepackage{subfigure}
\usepackage{cleveref}
\usepackage{multirow} 

\def\cm{\mathrm{cm}} 
\def\km{\mathrm{km}} 
\def\kpc{\mathrm{kpc}} 
\def\sec{\mathrm{s}} 
\def\TeV{\mathrm{TeV}} 
\def\GeV{\mathrm{GeV}} 

\makeatother

\allowdisplaybreaks 

\def\apj{Astrophys.\ J.\ }
\def\apjl{Astrophys.\ J.\ Lett.\ }

\def\ssr{Space\ Sci.\ Rev.\ }

\begin{document}

\title{A systematic study on the cosmic ray antiproton flux}
\author{Su-Jie Lin$^1$, Xiao-Jun Bi$^1$, Jie Feng$^2$, Peng-Fei Yin$^1$, Zhao-Huan Yu$^{1, 3}$}
\affiliation{$^1$Key Laboratory of Particle Astrophysics,
Institute of High Energy Physics, Chinese Academy of Sciences,
Beijing 100049, China}
\affiliation{$^2$School of Physics and Engineering, Sun Yat-Sen
University, Guangzhou 510275, P. R. China}
\affiliation{$^3$ARC Centre of Excellence for Particle Physics at the Terascale,
School of Physics, The University of Melbourne, Victoria 3010, Australia}

\begin{abstract}
Recently the AMS-02 collaboration has published the measurement of the cosmic
antiproton to proton ratio $\bar{p}/p$ and the $\bar{p}$ flux with a high
precision up to $\sim 450$~GeV.  In this work, we perform a systematic analysis
of the secondary antiproton flux generated by the cosmic ray interaction with
the interstellar gas.  The uncertainty of the prediction originates from the
cosmic ray propagation process and the hadronic interaction models.  Although
the cosmic ray propagation parameters have been well controlled by the AMS-02
$B/C$ ratio data for a specified model, different propagation models can not be
discriminated by the $B/C$ data.  The $\bar{p}$ flux is also calculated for
several hadronic interaction models, which are generally adopted by the cosmic
ray community. However, the results for different hadronic models do not
converge.  We find the EPOS LHC model, which seems to fit the collider data very
well, predicts a slightly lower $\bar{p}/p$ ratio than the AMS-02 data at the
high energy end.  Finally we derive the constraints on the dark matter
annihilation cross section from the AMS-02 $\bar{p}/p$ ratio for different
propagation and hadronic interaction models.
\end{abstract}


\maketitle

\section{Introduction}

Antimatter searches in cosmic rays (CRs) are especially important for
astrophysics and dark matter (DM) indirect detection.  In recent years, there
were great progresses in the measurements of CR antimatter particles. In
particular, the Alpha Magnetic Spectrometer (AMS-02) launched in 2011 has
provided unprecedentedly precise CR measurements. Based on the results from
AMS-02 \cite{AMSWebSite} and previous experiments, such as PAMELA
\cite{PAMELAWebSite} and Fermi-LAT\cite{FermiWebSite}, the properties of CR
antimatter particles have been extensively studied in a quantitative way in the
literature.

Recently, the AMS-02 collaboration has published the measurements of the
$\bar{p}/p$ ratio and the $\bar{p}$ flux up to $\sim 450$~GeV with a high
precision~\cite{Aguilar:2016kjl}.  The major challenge to interpret the AMS-02
data and search for potential DM signals is how to predict the secondary
antiproton flux precisely.  The secondary antiprotons are generated by collision
between the high energy cosmic rays and the interstellar gas. Its flux is
determined by the CR propagation process and its hadronic interaction with the
interstellar gas, both of which have relatively large uncertainties at present.
In this work, we perform a systematic analysis of the secondary antiproton flux
considering these possible uncertainties carefully.  Some relevant discussions
can also be found in Refs.  \cite{Giesen:2015ufa,Lin:2015taa,Jin:2015mka,Chen:2015cqa,Ibe:2015tma,Hamaguchi:2015wga,Kohri:2015mga,Kappl:2015bqa,Lu:2015pta,Feng:2016loc}

The CR propagation in the Galaxy involves many complicated effects, such as the
diffusion, energy loss, convection, and reacceleration. The propagation
parameters describing these effects can be determined by fitting to the
secondary-to-primary nuclei ratios, such as B/C and (Sc+Ti+V)/Fe, and
unstable-to-stable ratios of secondary nuclei, such as
$^{10}\mathrm{Be}/^9\mathrm{Be}$ and $^{26}\mathrm{Al}/^{27}\mathrm{Al}$
~\cite{Strong:1998pw,1990cup..book.....G,DiBernardo:2009ku,Maurin:2001sj}, where
the secondary nuclei are generated by the primary nucleus collision with the
interstellar matter when they are propagating.  However, the current
measurements of these ratios remain not sufficient to distinguish different
propagation models. Therefore the propagation is one of the main sources of
uncertainty when predicting the secondary antiproton flux. Another important
uncertainty of propagation comes from the solar modulation
\cite{Gleeson:1968zza}.  This effect significantly affects low energy CR
spectra, but is difficult to precisely quantify.

In this work, the GALPROP package is used to solve the CR propagation
equation~\cite{Strong:1998pw,Moskalenko:1997gh}. We adopt three kinds of
propagation models, namely the diffusion-reacceleration (DR) model, the modified
diffusion-reacceleration (DR-2) model, and the diffusion-convection (DC) model.
Propagation parameters are determined by fitting the available B/C data with a
Markov Chain Monte Carlo (MCMC, \cite{Lewis:2002ah}) algorithm
~\cite{Liu:2009sq,Liu:2011re,Yuan:2013eja,Lin:2014vja}, which is powerful for
surveying high dimensional parameter space.

The hadronic interaction between CR particles and interstellar gas is another
important source of uncertainty for calculating the antiproton flux. Although
Quantum Chromodynamics (QCD) is well-established as the theory of strong
interaction and has been confirmed at collider experiments, only the processes
with large momentum transfers can be predicted by the perturbative calculation
from the first principle. It is not possible to calculate the forward scattering
processes with multiparticle production by QCD, such as the CR interaction with
interstellar gas considered here.  In order to calculate such hadronic
processes, some simplified assumptions and phenomenological/empirical
parametrization are needed.

In the cosmic ray study, it has been a long time difficulty to deal with the
hadronic interaction. Many phenomenological forms have been proposed in the
literature to describe the hadronic interaction.  Unfortunately, there is no
consensus on the empirical hadronic model so far. In this work, we discuss the
impact of hadronic models on the antiproton production. We find that the
hadronic interaction induces the largest uncertainty in the prediction of the
antiproton flux.

Based on the predicted $\bar{p}$ flux and the AMS-02 measurement, we derive
constraints on the DM annihilation cross section in different hadronic
interaction and CR prorogation models.

This paper is organized as follows. In Sec.~II, we describe the CR propagation
processes. In Sec.~III, we introduce the hadronic interaction models and compare
the prediction with accelerator data.  The secondary antiproton flux and
$\bar{p}/p$ ratio prediction are presented in Sec. IV.  In Sec.~V, we
investigate the implications for DM annihilation from the AMS-02 $\bar{p}/p$
measurement and the CRs prediction.  Finally, we give the summary and
discussions in Sec.~V.

\section{Propagation of Galactic cosmic rays}

After accelerated in sources, Galactic CRs are injected and diffuse in the
interstellar space and suffer from several propagation effects before arriving
at the Earth. A conventional assumption is that CRs propagate in a cylindrical
halo with a half height $z_h$, beyond which CRs escape freely.  The propagation
equation can be expressed as~\cite{Strong:2007nh}
\begin{eqnarray}
\frac{\partial \psi}{\partial t} &=& Q(\mathbf{x}, p) + \nabla \cdot ( D_{xx}\nabla\psi - \mathbf{V}_{c}\psi )
+ \frac{\partial}{\partial p}\left[p^2D_{pp}\frac{\partial}{\partial p}\left(\frac{\psi}{p^2}\right)\right]
\nonumber\\
&& - \frac{\partial}{\partial p}\left[ \dot{p}\psi - \frac{p}{3}(\nabla\cdot\mathbf{V}_c)\psi \right]
- \frac{\psi}{\tau_f} - \frac{\psi}{\tau_r},
\label{propagation_equation}
\end{eqnarray}
where $Q(\mathbf{x}, p)$ is the CR source term, $\psi=\psi(\mathbf{x},p,t)$ is
the CR density per momentum interval, $\dot{p}\equiv \mathrm{d}p/\mathrm{d}t$ is
the momentum loss rate, and the time scales $\tau_f$ and $\tau_r$ characterize
fragmentation processes and radioactive decays, respectively.

It is generally believed that CR particles are accelerated in supernova remnants
(SNRs).  Thus the spatial distribution of the CR sources follows the SNR
distribution as
\begin{equation}
f(r,z) = \left( \frac{r}{r_\odot} \right)^a \exp\left[ -\frac{b(r-r_\odot)}{r_\odot} \right]\exp\left( -\frac{\left|z\right|}{z_s} \right),
\label{spatial_distribution}
\end{equation}
where $r_\odot=8.5$~kpc is the distance from the Sun to the Galactic Center, and
$z_s\simeq 0.2$~kpc is the characteristic height of the Galactic disk.  We adopt
$a=1.25$ and $b=3.56$~\cite{Trotta:2010mx} in the calculation.  The injection
spectra are assumed to be broken power-law functions with respect to the
rigidity.

The spatial diffusion coefficient $D_{xx}$ can be parametrized as
\cite{Maurin:2010zp}
\begin{equation}
D_{xx} = D_0\beta^\eta \left( R/R_0 \right)^{\delta},
\end{equation}
where $R\equiv pc/Ze$ is the rigidity, $\beta$ is the CR particle velocity in
units of the light speed $c$, $R_0$ is the reference rigidity and $D_0$ is a
normalization parameter. Although the slop of diffusion coefficient $\delta$ is
predicted to be $\delta=1/3$ for a Kolmogorov spectrum of interstellar
turbulence, or $\delta=1/2$ for a Kraichnan cascade, this parameter is usually
treated as a free parameter when explaining data. The factor of $\beta^\eta$
denotes the effect that the diffusion coefficient could be altered at low
velocities due to the turbulence dissipation.  We take $\eta=1$ as the standard
diffusion plus reacceleration (DR) case taken in GALPROP.  We refer the case
with $\eta=-0.4$ as the DR-2 model, which can improve the secondary fitting as
discussed in \cite{DiBernardo:2010is}.

The CR reacceleration process due to collisions on interstellar random weak
hydrodynamic waves can be described by the diffusion in momentum space with a
coefficient $D_{pp}$, which is related with $D_{xx}$
by~\cite{1994ApJ...431..705S}
\begin{equation}
D_{pp} D_{xx}=\frac{4p^2v^2_{A}}{3\delta(4-\delta^2)(4-\delta)\omega }\ \ ,
\label{reacceleration}
\end{equation}
where $v_{A}$ is the Alfv\'{e}n velocity and $\omega$ is the ratio of the
magnetohydrodynamic wave energy density to the magnetic field energy density.
Since $D_{pp}$ is proportional to $v^2_A/\omega$, we can set $\omega=1$ and just
use $v_A$ to characterize the reacceleration effect.  The convection velocity
$\mathbf{V}_c$ is usually assumed to linearly depend on the distance away from
the Galactic disk and the convection effect can be described by the quantity
$\mathrm{d}V_c/\mathrm{d}z$.  Therefore, the major propagation parameters
involve $D_0$, $\delta$, $R_0$, $\eta$, $v_A$, $\mathrm{d}V_c/\mathrm{d}z$, and
$z_h$.

When CRs propagate in the solar system, their spectra with $R\lesssim 20$~GV are
significantly affected by solar winds.  This is the solar modulation effect,
which depends on solar activities and varies with the solar cycle.  It can be
described by the force field approximation~\cite{Gleeson:1968zza} with only one
parameter, i.e., the solar modulation potential $\phi$.

Due to fragmentation and radioactive decays, secondary CR particles are produced
in propagation processes.  As a result, secondary-to-primary ratios and
unstable-to-stable secondary ratios, such as B/C and
$^{10}\mathrm{Be}/^9\mathrm{Be}$, are sensitive to propagation parameters, but
almost independent of primary injection spectra.  Thus the measurements of these
ratios are useful for determining the propagation parameters (see e.g.
Refs.~\cite{Putze:2010zn,Trotta:2010mx}).

Note that there are degeneracies between the propagation models with the
reacceleration process and those with the convection effect. Current
measurements are not sufficient to distinguish these models.  Therefore, we
separately consider the DR, DR-2, and DC models when calculating the antiproton
flux.  $R_0$ is taken to be 4~GV in the DR and DR-2 models.  In the DC model,
however, $R_0$ is set to be a free parameter and $\delta=0$ is imposed for
$R<R_0$.

\section{hadronic interaction models}

Although QCD has been proved to be a very successful theory of the strong
interaction, we do not have a realistic technique to calculate the bulk of
multiparticle production in hadronic interactions from the first principle.
Many phenomenological and empirical models have been developed to explain the
collider and comic ray data.  These models include a large number of parameters
that encode the fundamental physics and phenomenological descriptions of the
fragmentation process.

These hadronic models could be classified into three categories according to
their purposes.  HERWIG\cite{Corcella:2000bw}, PYTHIA\cite{Sjostrand:2006za} and
SHERPA\cite{Gleisberg:2008ta} are generally adopted in high energy physics (HEP)
studies, and focus on hard-scattering processes at accelerator experiments.  In
contrast, QGSJET01\cite{Kalmykov:1997te}, QGSJET
II\cite{Ostapchenko:2004ss,Ostapchenko:2005nj} and
SIBYLL\cite{Engel:1992vf,Fletcher:1994bd,Ahn:2009wx} are designed for simulating
the extensive air showers caused by the high energy CR particles. Therefore,
these models focus on the bulk production of soft particles with low transverse
momenta, and emphasize on providing a reasonable extrapolation for higher energy
and wider phase-space regions.
PHOJET\cite{Engel:1994vs,Engel:1995yda,Engel:1995sb}, DPMJET\cite{Bopp:2005cr}
and EPOS\cite{Werner:2005jf,Pierog:2013ria} fall in between these two
categories.  These models can also approach the HEP models regarding the hard
process, and are adjusted to well reproduce the low energy accelerator data.

The HEP models listed above are not considered in our calculation, as they are
not suitable for dealing with CRs interaction with the interstellar gas. As the
hadronic models proposed for the CR interaction mainly focus on the high energy
region, the description of the interaction for CRs below 100 GeV may be not
precise. Fortunately, in recent years many new data sets measured at LHC and
fixed-target experiments are available.  New versions of QGSJET II-04
\cite{Ostapchenko:2010vb} and EPOS LHC\cite{Pierog:2013ria} have been released
whose model parameters are tuned to fit these data
\cite{Chatrchyan:2012qb,Csorgo:2012dm,Aamodt:2010pp,Aamodt:2011zj,Aad:2010ac}.
The EPOS LHC model fits the available accelerator antiproton data in all energy
regions.  Recently the QGSJET II-04 has been slightly modified to better
reproduce the accelerator antiproton data at low energies, referred to as QGSJET
II-04m here \cite{Kachelriess:2015wpa}. In the following, we use EPOS LHC and
QGSJET II-04m to calculate the CR antiproton flux.  Another version of EPOS,
namely EPOS 1.99\cite{Pierog:2009zt}, is also considered in the calculation,
since this version has been more carefully tuned to fit the low energy
accelerator data, such as the measurements of NA49 and
NA61\cite{Grebieszkow:2009th}.

For comparison, we also consider the default hadronic model embedded in GALPROP
\cite{Moskalenko:1998id}, which is referred to as Tan \& Ng + BP01. In this
model, a parametrization given by Tan \& Ng \cite{1983JPhG....9.1289T} and a
code from Barashenkov \& Polanski are combined to estimate the $\bar{p}$
spectrum.  Other models for high energy air showers (QGSJET01, QGSJETII and
SIBYLL) are not used to predict the CR antiproton flux, as they do not aim at
describing the low energy CR interactions considered here.

\begin{figure}[!htp]
  \centering
  \includegraphics[width=0.5\textwidth]{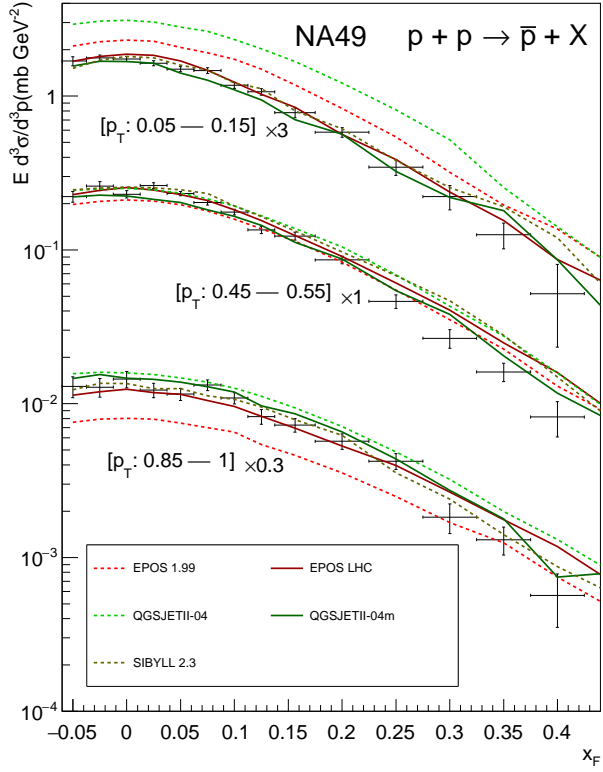}
  \caption{ The differential cross section of $E \, d^3 \sigma /d^3 p$ for the $p+p\rightarrow\bar{p}+X$ process compared with the NA49 measurement
\cite{Anticic:2009wd}, with variables given in the center-of-mass system. Three groups of curves from the top to bottom correspond to the $p_T$ ranges of $[0.05-0.15]\GeV$, $[0.45-0.55]\GeV$ and $[0.85-1]\GeV$, with a rescale factor of 3, 1 and 0.3, respectively.
$x_{F}$ is defined as $x_{F}\equiv 2 p_{z}/\sqrt{s}$.
  }
  \label{fig:compare_generator_na49_pt}
\end{figure}

\begin{figure}[!htp]
  \centering
  \includegraphics[width=0.5\textwidth]{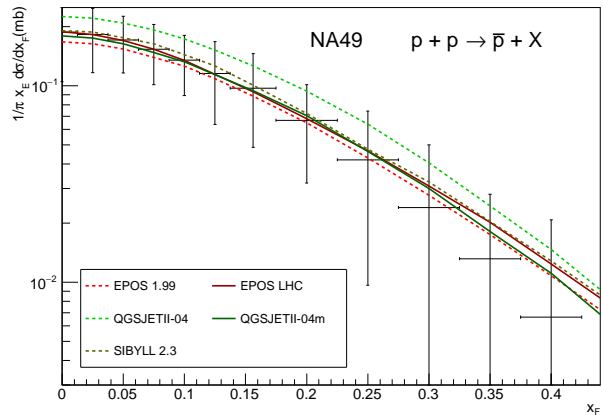}
  \caption{The differential cross section of
    $1/\pi\,x_{E}\mathrm{d}\sigma/\mathrm{d}x_F$ for the
    $p+p\rightarrow\bar{p}+X$.  $x_{E}$ is defined as $x_{E}\equiv 2E_{\bar{p}}
    /\sqrt{s}$ in the center-of-mass system.
}
  \label{fig:compare_generator_na49_total}
\end{figure}

\begin{figure}[!htp]
  \centering
  \includegraphics[width=0.5\textwidth]{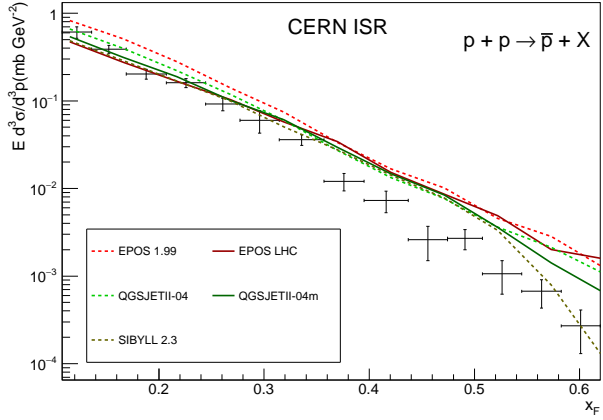}
  \caption{ The differential cross section of $E \, d^3 \sigma /d^3 p$
for the $p+p\rightarrow\bar{p}+X$ process
compared with the CERN ISR experiment\cite{Albrow:1973kj} with $\sqrt{s}=53\GeV$.
}
  \label{fig:compare_generator_ISR}
\end{figure}

\begin{figure}[!htp]
  \centering
  \includegraphics[width=0.5\textwidth]{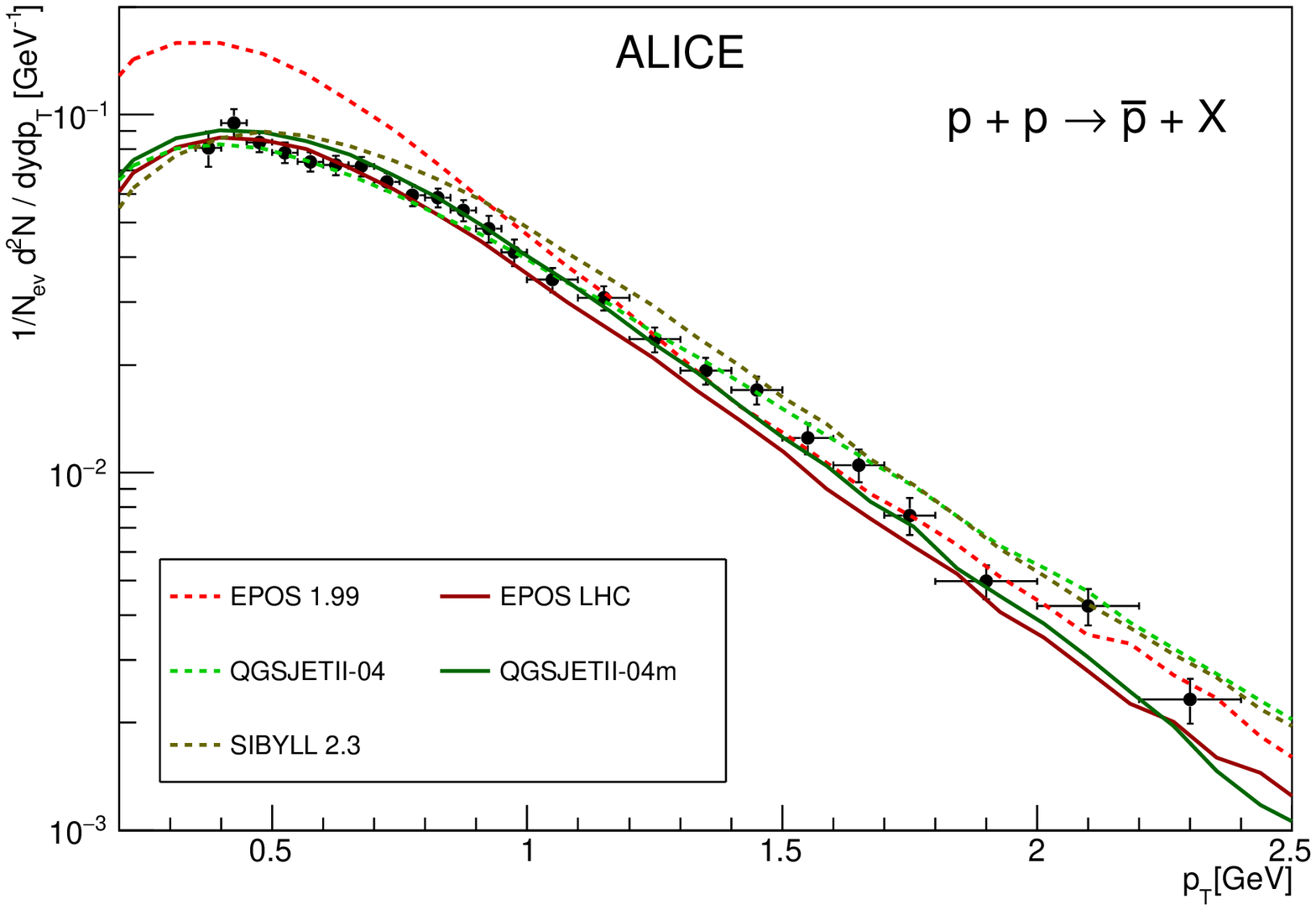}
  \caption{The antiproton spectrum $1/N d^2 N/dy dp$ for the $p+p\rightarrow\bar{p}+X$ process as a function of the transverse momentum $p_T$ in the center-of-mass system with $\sqrt{s}=900\GeV$. The rapidity $y$ is cut with the condition $\left|y\right|<0.5$. Also shown are the data from the ALICE experiment \cite{Aamodt:2011zj}.}
  \label{fig:compare_generator_alice}
\end{figure}

In the following we compare the expectations of different hadronic models with
the accelerator measurements of the process $pp\rightarrow \bar{p} X$.  In Figs.
\ref{fig:compare_generator_na49_pt} and \ref{fig:compare_generator_na49_total}
we show the cross section of $pp\to \bar{p}+X$ as a function of
$x_F\equiv2p_z/\sqrt{s}$ in the center-of-mass system for specified transverse
momentum $p_T$ and integrated cross section over $p_T$ for different hadronic
models respectively. The expectation is compared with the NA49 data
\cite{Anticic:2009wd,Baatar:2012fua}, which is a fixed target experiment with
the incoming beam momentum of $158\GeV/c$.  In Fig.
\ref{fig:compare_generator_na49_pt}, we consider three representative ranges of
$p_T$ standing for the cases of low, medium and high $p_T$.  As can be seen that
all the models reproduce the cross section integrated over $p_T$ well, but only
QGSJET II-04m, EPOS LHC and SIBYLL could fit the differential cross section for
the specified $p_T$.

For a higher energy region, we compare the model prediction with the data from
the CERN ISR experiment \cite{Albrow:1973kj} with the center-of-mass energy of
$\sqrt{s}=53\GeV$.  Such energy is relevant to the secondary antiprotons with
energies of hundreds of $\GeV$.  In Fig. \ref{fig:compare_generator_ISR}, we
show the expectations of differential cross section for $pp\to\bar{p}X$ and
compare with the ISR data.  As can be seen that all the selected models
reasonably reproduce the data in low $x_F$ region, while slightly overestimate
the antiproton in the high $x_F$ region.

Finally, we compare the expectations with the measurement of the ALICE
experiment at $\sqrt{s}=900\GeV$\cite{Aamodt:2011zj}.  The antiprotons are for
the rapidity $\left|y\right|<0.5$.  We find that all the selected hadronic
models could give a good fit to the ALICE results except EPOS 1.99.  Thus we
conclude that the EPOS LHC and QGSJET II-04m models are suitable to predict the
CR antiproton as they can interpret all the accelerator antiproton data well.
Note that the center-of-mass energy $900\GeV$ corresponding to a CR proton
energy of $430\TeV$ is far beyond the energy region relevant to the AMS-02
antiproton data. Therefore the EPOS 1.99 model is also adopted to calculate the
CR antiproton flux, as it can interpret the low energy accelerator data well.

\section{Astrophysical prediction for the $\bar{p}/p$ ratio}

In this section, we calculate the secondary antiproton flux generated by CRs
when they propagate in the Galaxy, and study the uncertainties from the
propagation and hadronic interaction models.

We use the numerical tool GALPROP~\cite{Moskalenko:1997gh,Strong:1998pw} to
solve the propagation equation. The propagation parameters are determined by
fitting the B/C and $^{10}$Be/$^9$Be data.  In order to improve the fitting
efficiency, the MCMC method is employed to derive the posterior probability
distributions of propagation parameters.  By fitting the B/C ratio data from
AMS-02~\cite{AMS_ICRC13} and ACE~\cite{Mewaldt:2000zz}, and the
$^{10}\mathrm{Be}/^9\mathrm{Be}$ ratio data
\cite{2001ApJ...563..768Y,1977ApJ...212..262H,1978ApJ...226..355B,1979ICRC....1..389W,1981ICRC....2...72G,1988SSRv...46..205S,Hams:2004rz,1998ApJ...501L..59C,1999ICRC....3...41L},
we obtain the mean values and $1\sigma$ errors of propagation parameters for the
three propagation models, as shown in Table~\ref{tab:BtoC_fit}.  Here the proton
injection spectrum is determined by fitting the AMS-02~\cite{AMS_ICRC13} and
CREAM~\cite{Ahn:2010gv} proton data. The proton spectrum hardening at $\sim 200$
GeV can enhance the $\bar{p}/p$ ratio at the high energy end.

\begin{table*}
\centering
\setlength\tabcolsep{0.4em}
\caption{Mean values and $1\sigma$ uncertainties of the propagation parameters derived through fitting the data of B/C  and $^{10}\mathrm{Be}/^9\mathrm{Be}$ ratios in three propagation models.}
\label{tab:BtoC_fit}
\begin{tabular}{cccc}
\hline\hline
 & DR & DR-2 & DC \\
\hline
$D_0$ ($10^{28}~\cm^2~\sec^{-1}$) & $6.58\pm 1.27$ & $3.59\pm 0.88$ & $1.95\pm 0.50$ \\
$\delta$ & $0.333\pm 0.011$ & $0.423\pm 0.017$ & $0.510\pm 0.034$ \\
$R_0$ (GV) &  4 & 4 & $4.71\pm 0.80$ \\
$v_A$ ($\km~\sec^{-1}$) & $37.8\pm 2.7$ & $22.6\pm 3.1$ & / \\
$\mathrm{d}V_c/\mathrm{d}z$ ($\km~\sec^{-1}~\kpc^{-1}$) & / & / & $4.2\pm 3.2$ \\
$z_h$ (kpc) & $4.7\pm 1.0$ & $3.5\pm 0.8$ & $2.5\pm 0.7$ \\
$\phi_\mathrm{B/C}$ (MV) & $326\pm 36$ & $334\pm 37$ & $182\pm 25$ \\
\hline\hline
\end{tabular}
\end{table*}

\begin{figure}[!htp]
  \centering
  \includegraphics[width=0.7\textwidth]{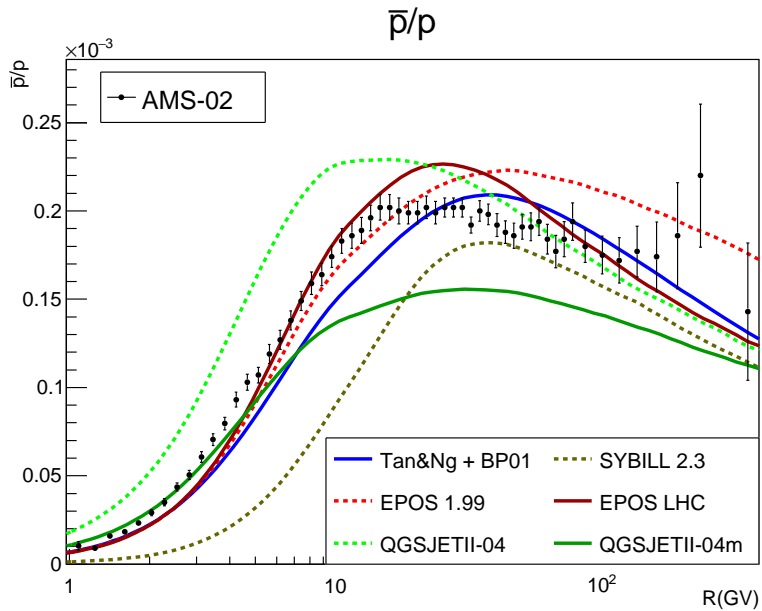}
  \caption{The $\bar{p}/p$ ratio expected from different hadronic models in the DR propagation model, in comparison with the AMS-02 measurement~\cite{Aguilar:2016kjl}. The propagation parameters used here give the best-fit to the B/C ratio, adopted from Ref.~\cite{Lin:2014vja}: $D_0=6.58\times 10^{28}\,\mathrm{cm^2\,s^{-1}}$, $\delta=0.33$, $v_A=37.8\,\mathrm{km\,s^{-1}}$, $z_h=4.7\,\mathrm{kpc}$, $\nu_1=1.81$, $\nu_2=2.40$. The solar modulation potential is set to $800~\mathrm{MV}$.}
  \label{fig:compare_generator}
\end{figure}

In Fig.~\ref{fig:compare_generator} we illustrate the $\bar{p}/p$ ratio in the
DR propagation model for different hadronic interaction models.  The QGSJET
II-04 overestimates the low energy antiprotons.  This is not surprising since it
does not reproduce the low energy accelerator data well as shown in Fig.
\ref{fig:compare_generator_na49_pt}.  The fitting is improved in the modified
QGSJET II-04 \cite{Kachelriess:2015wpa}. The SYBILL underestimates the low
energy antiprotons greatly.  The reason is that SYBILLL does not extend to the
cases with the incoming particles energy below   $\sim60\GeV$. We do not
consider the two interaction models later.

However, even the other four hadronic models provide diverse predictions.  Note
that the propagation parameters have been adjusted to fit the B/C data.
Therefore the discrepancies between the $\bar{p}/p$ predictions and data may be
due to the uncertainties in the nuclear reaction cross sections for
$\mathrm{C},\mathrm{O} \to \mathrm{B}$ or in the hadronic interactions for
$pp\to \bar{p}X$. The uncertainties may also come from the heavy element
interactions $pA (AA)\to \bar{p}X$.  In order to include such effects, we
introduce a free factor $c_{\bar{p}}$ to slightly rescale the antiproton flux in
order to fit the AMS-02 data.  Therefore we present the prediction of $\bar{p}$
flux and $\bar{p}/p$ ratio, which is ``calibrated'' by the AMS-02 data,  in the
following, since the AMS-02 collaboration provides very precise measurements and
well-controlled systematic errors.

\begin{figure}[!htp]
  \centering
  \includegraphics[width=0.4\textwidth]{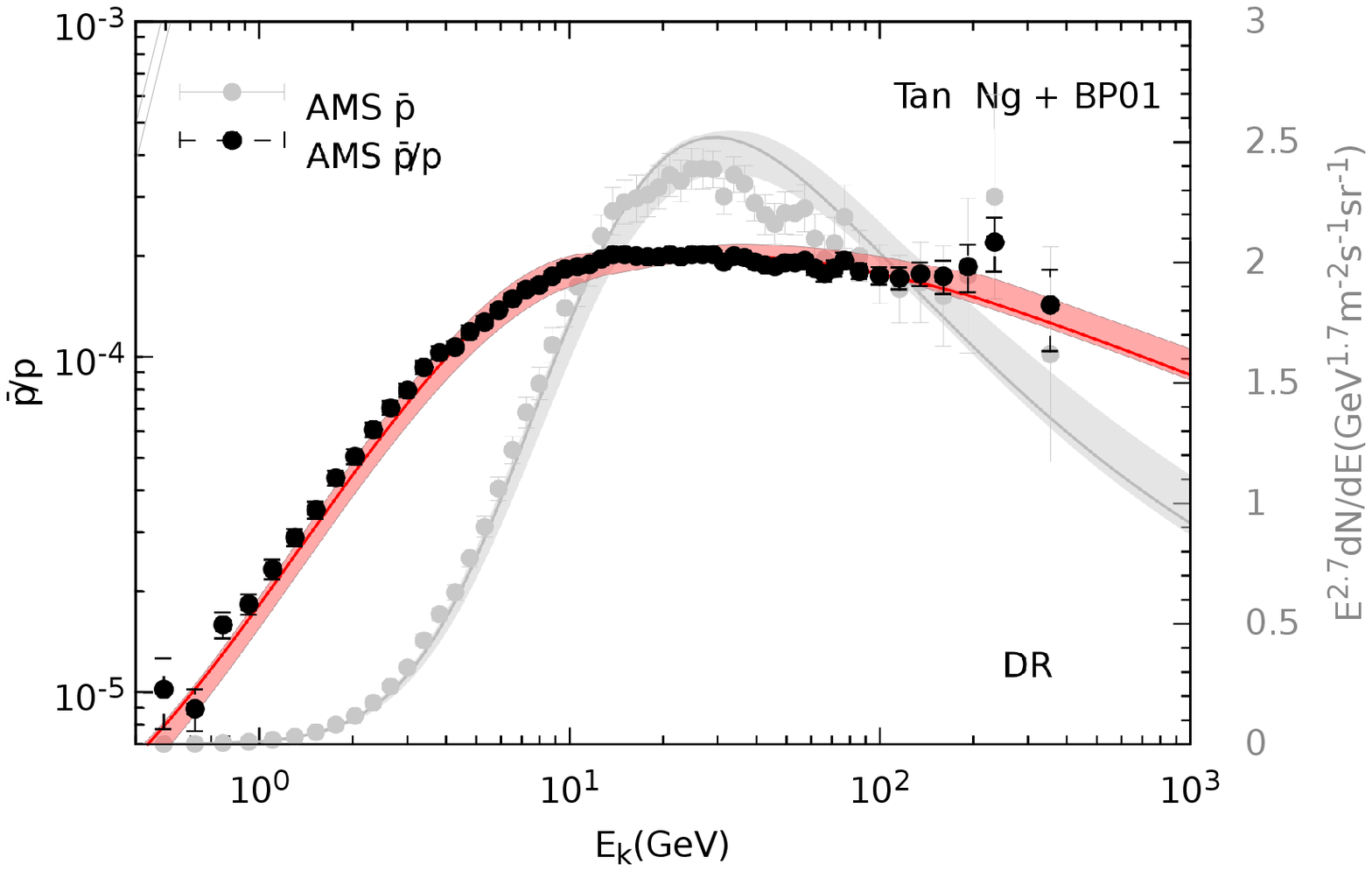}
  \includegraphics[width=0.4\textwidth]{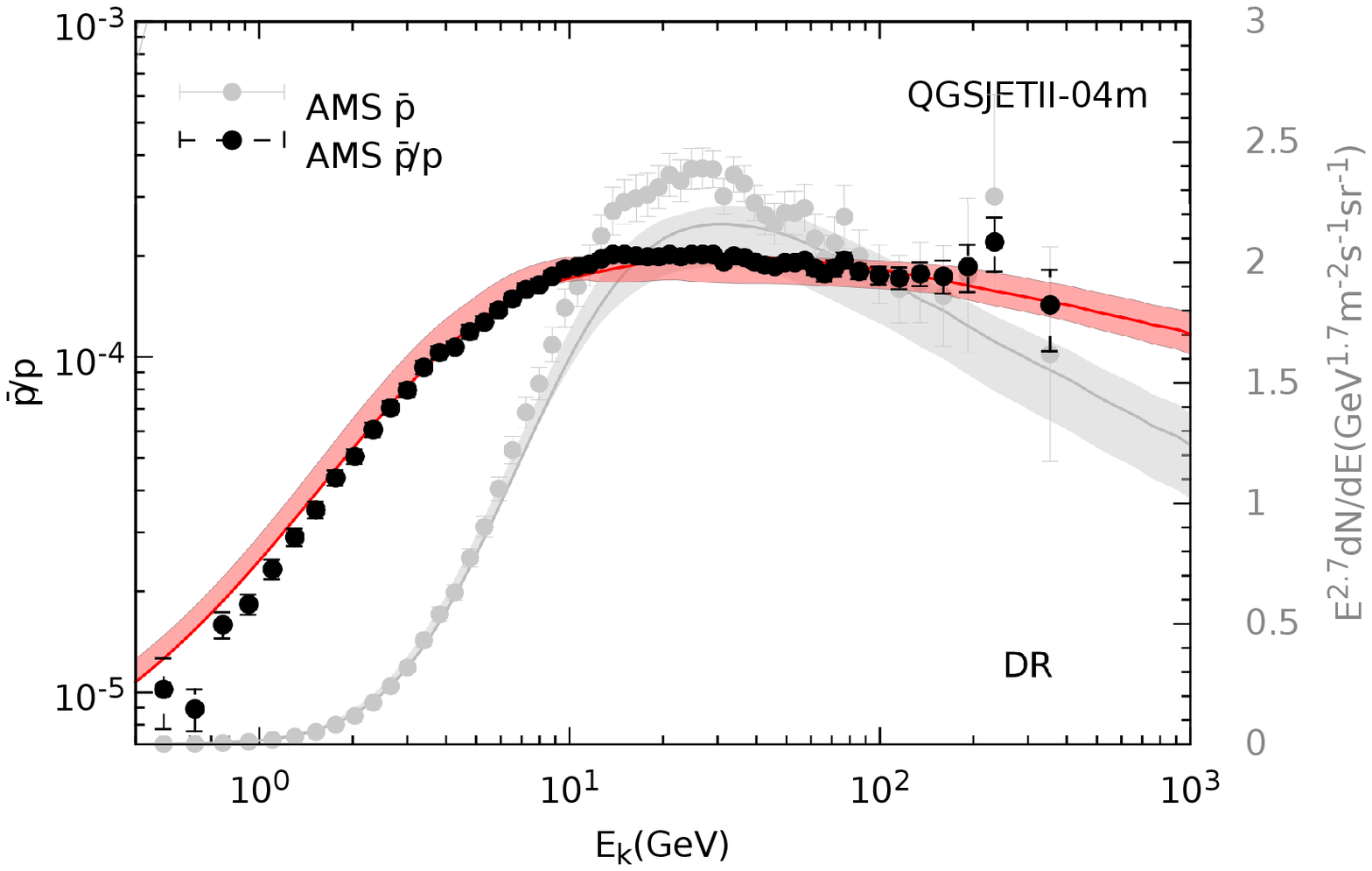}\\
  \includegraphics[width=0.4\textwidth]{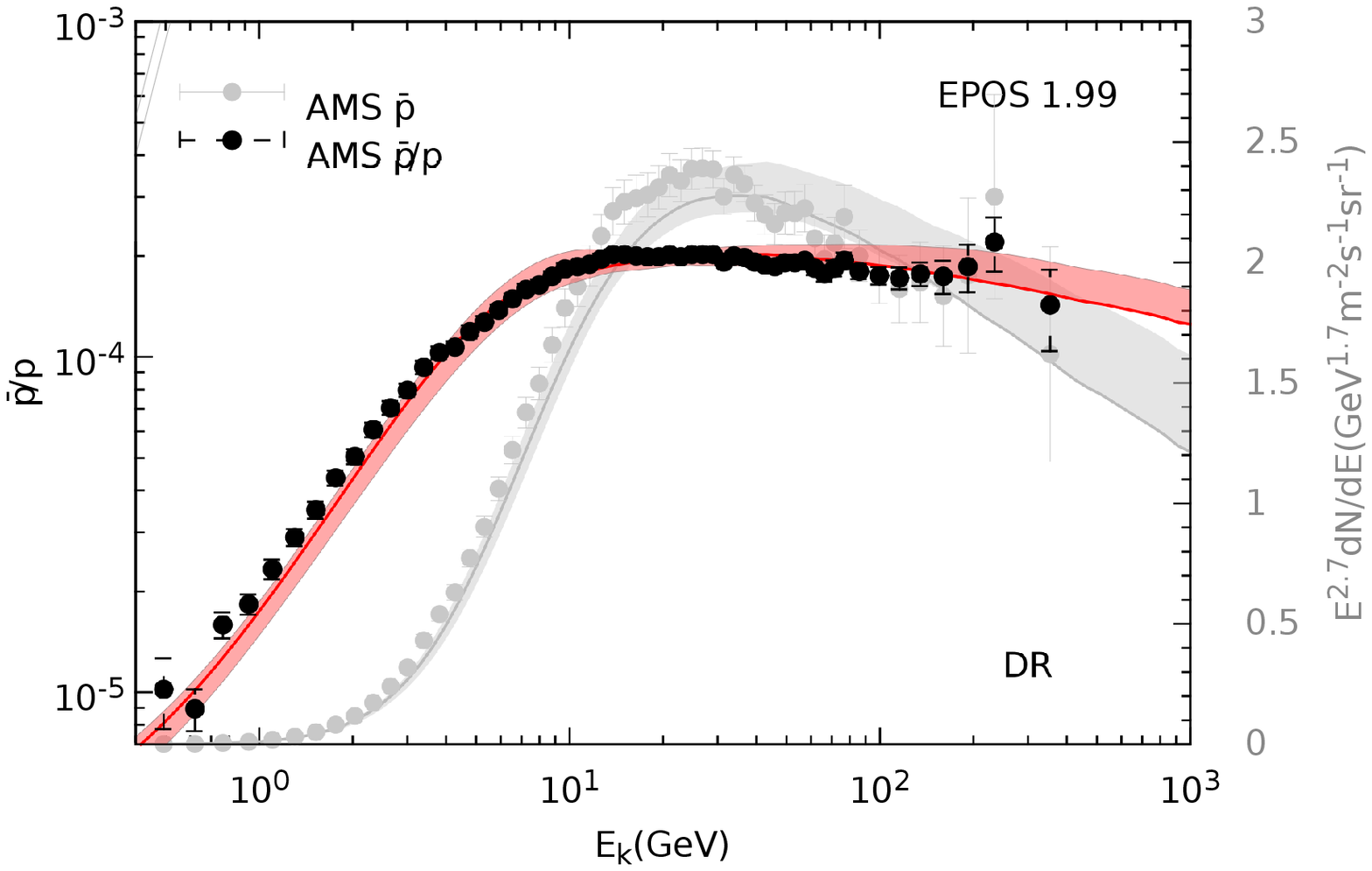}
  \includegraphics[width=0.4\textwidth]{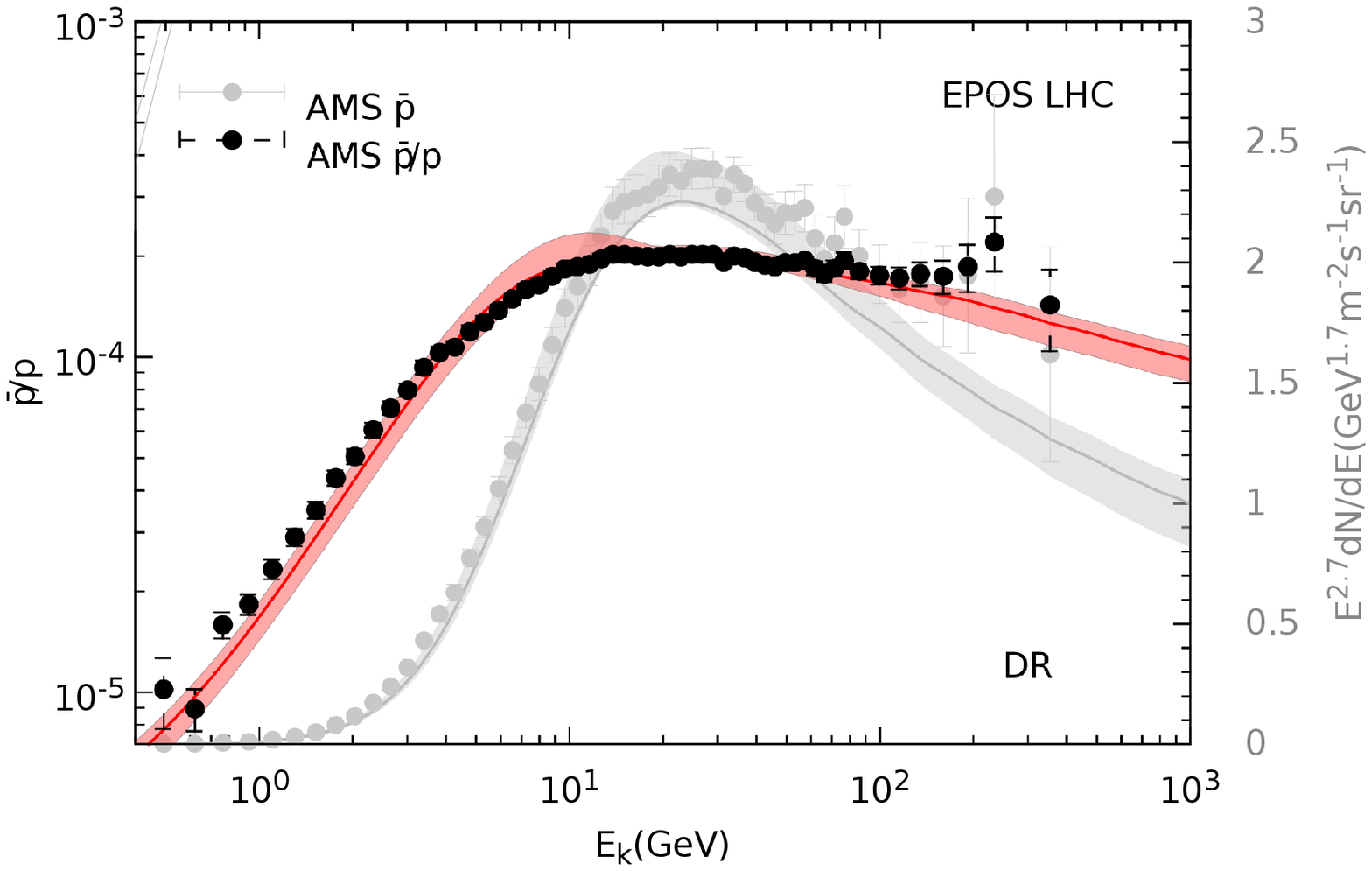}\\
  \caption{The $\bar{p}/p$ ratio (colored) and $\bar{p}$ flux (gray) for
  different hadronic models in the DR model, comparing with the AMS-02
  measurement~\cite{Aguilar:2016kjl}.  The bands indicate the propagation
  uncertainty in 95\% C.L., while the line is the best-fit case to the
  $\bar{p}/p$ ratio. We have tuned the $\bar{p}/p$ ratio with suitable rescaling
  and solar modulations to fit the data, and the $\bar{p}$ flux prediction
corresponds to this fit.}
  \label{fig:background_DR_cpbar}
\end{figure}

\begin{figure}[!htp]
  \centering
  \includegraphics[width=0.4\textwidth]{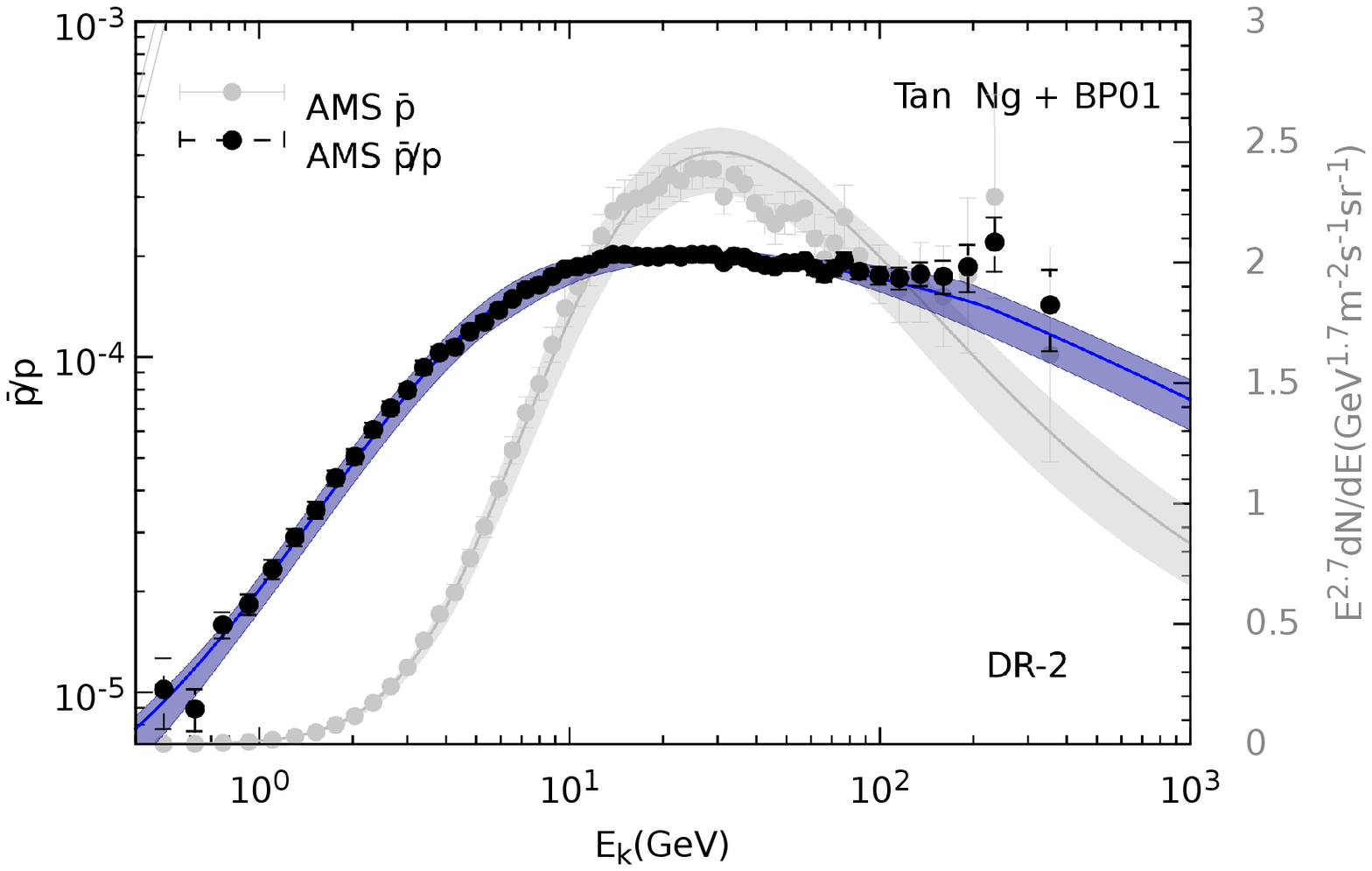}
  \includegraphics[width=0.4\textwidth]{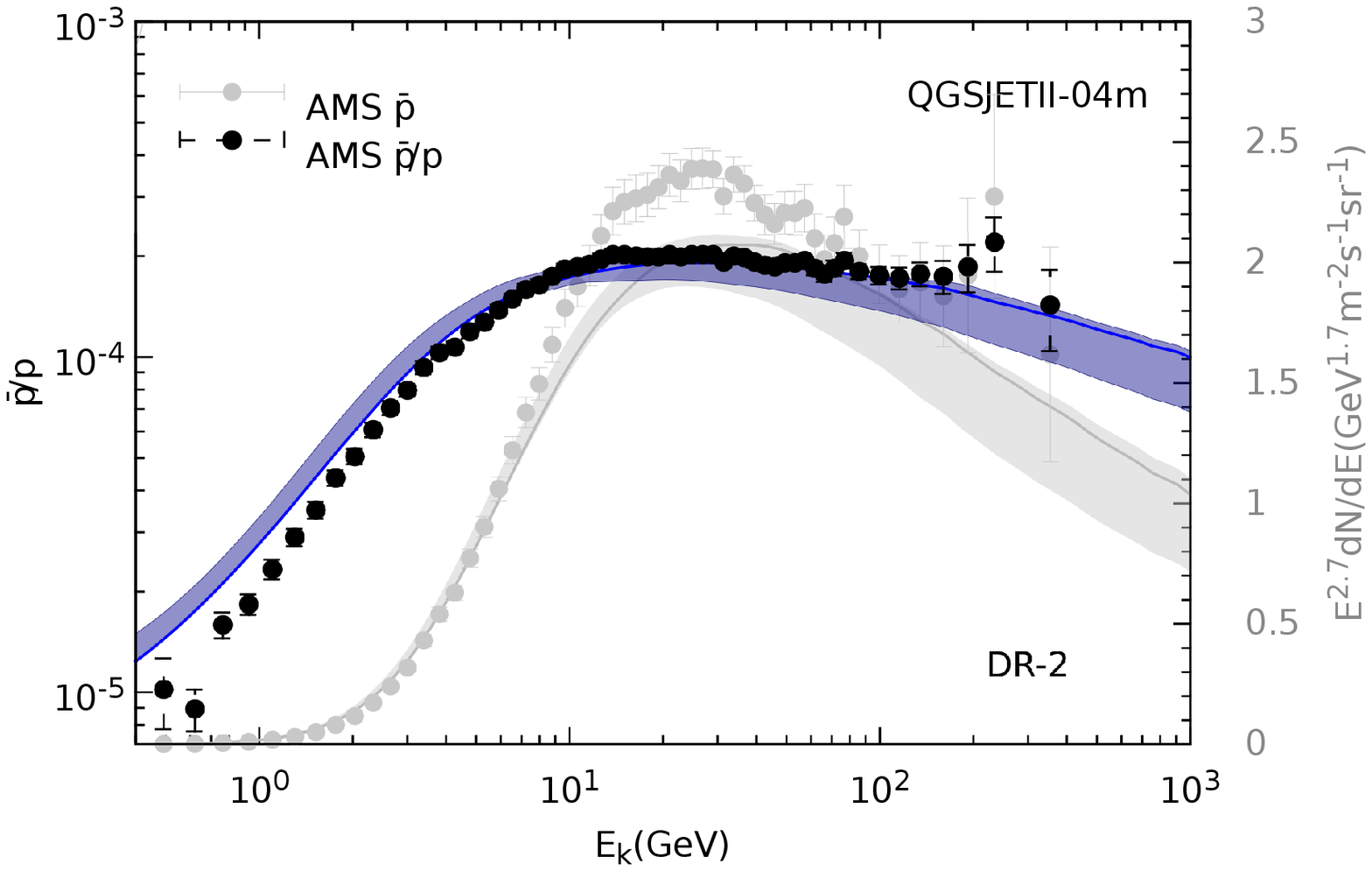}\\
  \includegraphics[width=0.4\textwidth]{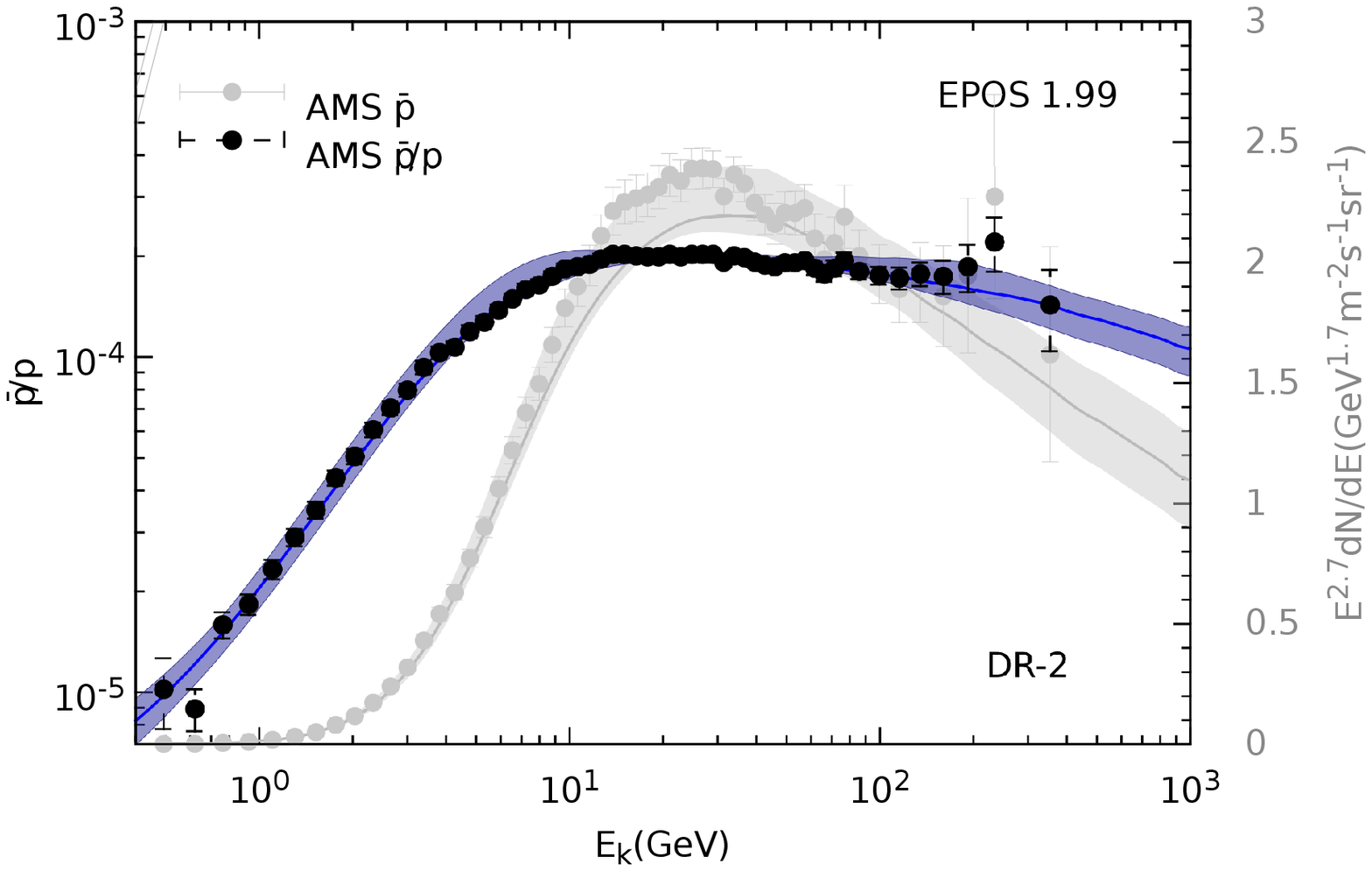}
  \includegraphics[width=0.4\textwidth]{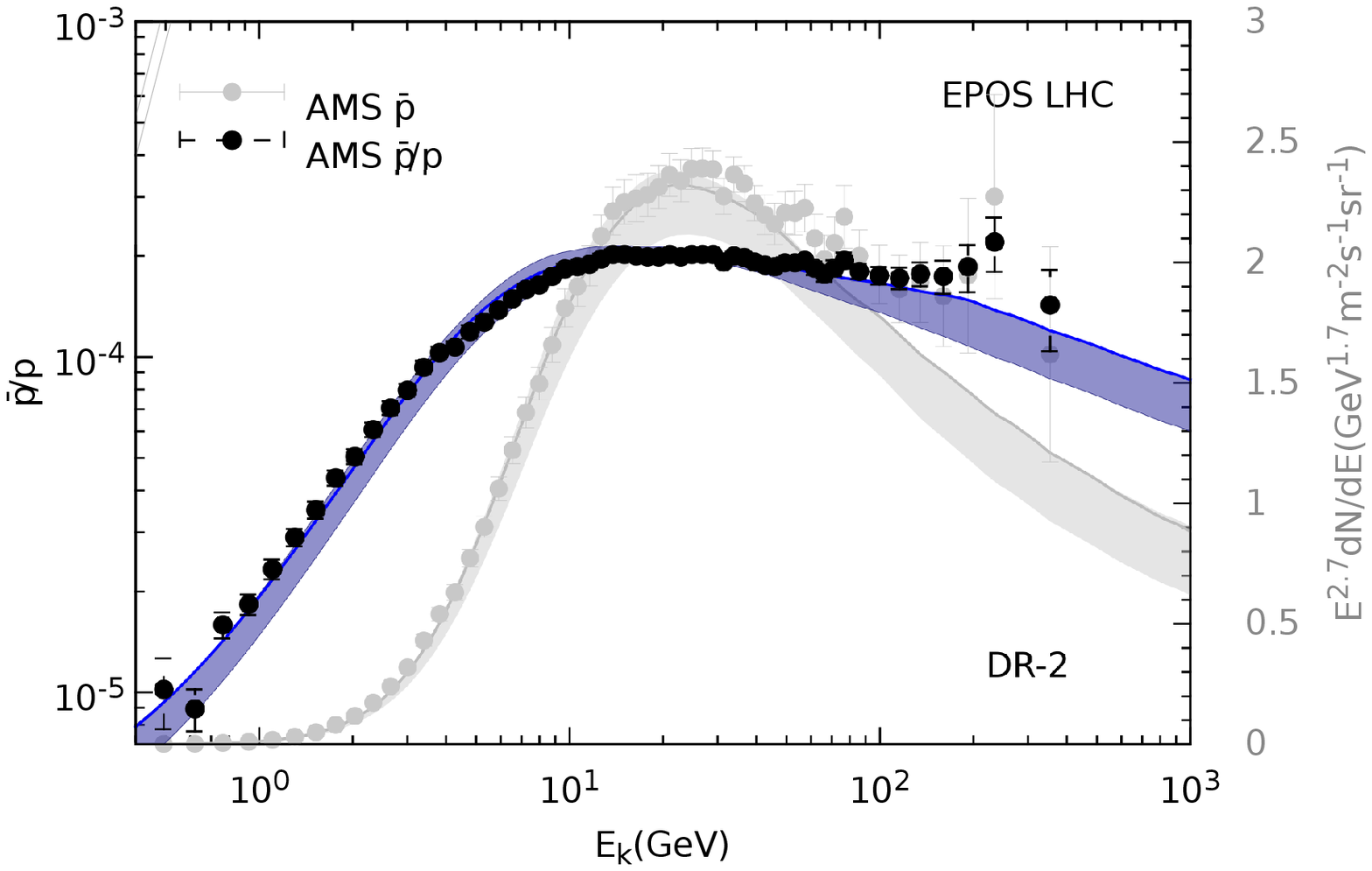}\\
  \caption{The same as \ref{fig:background_DR_cpbar} but for the DR-2 model.}
  \label{fig:background_DR2_cpbar}
\end{figure}

\begin{figure}[!htp]
  \centering
  \includegraphics[width=0.4\textwidth]{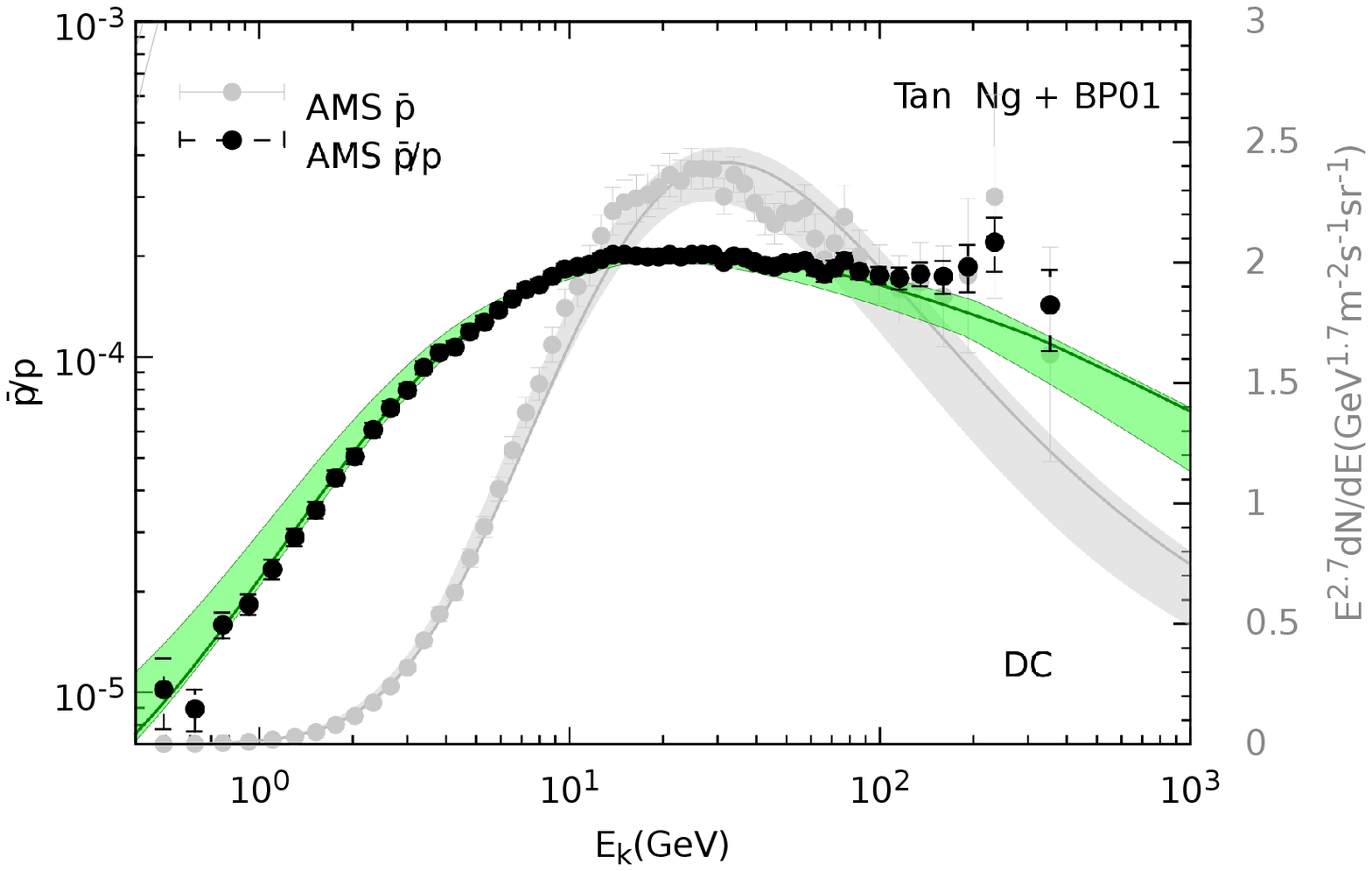}
  \includegraphics[width=0.4\textwidth]{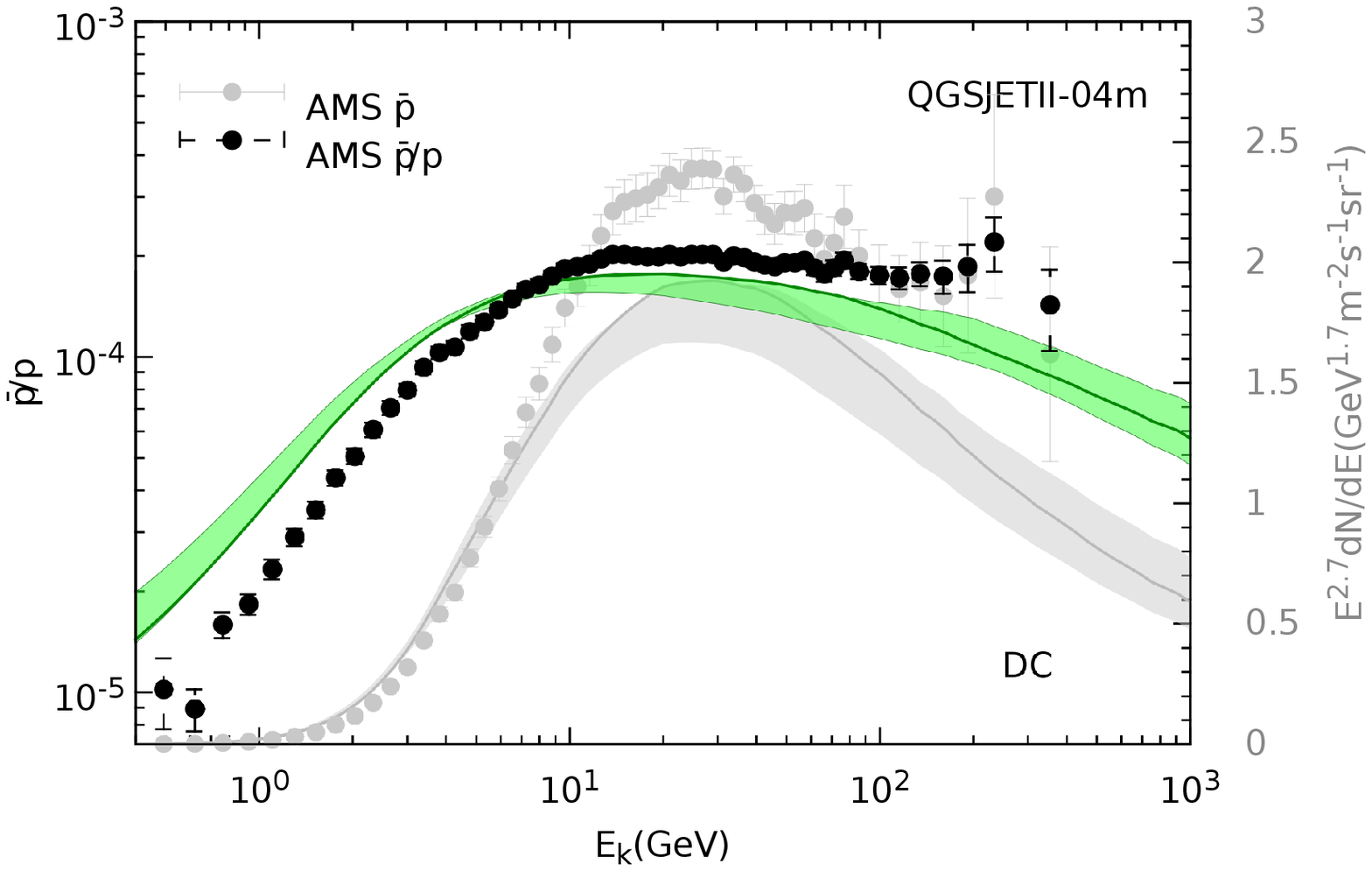}\\
  \includegraphics[width=0.4\textwidth]{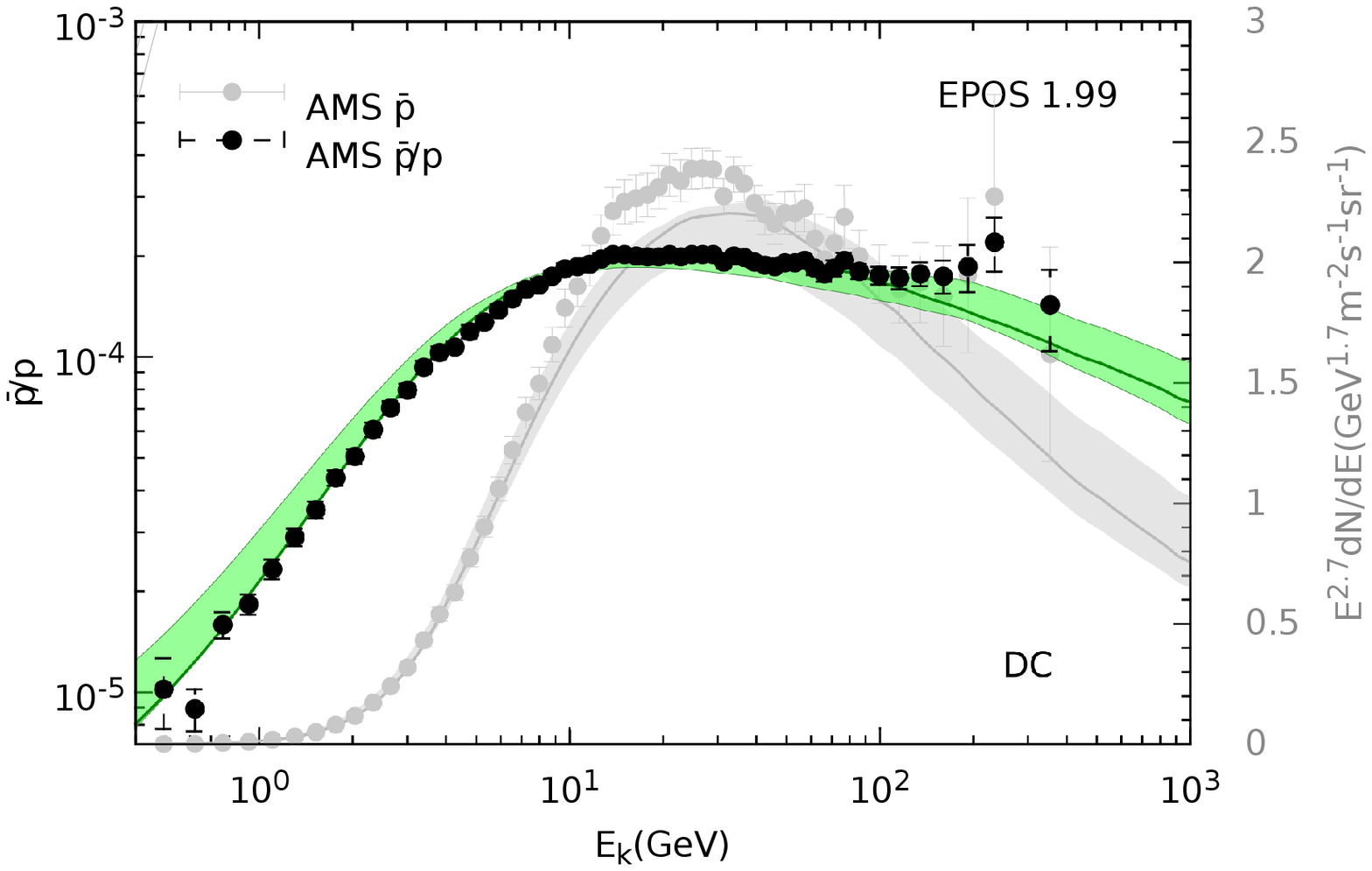}
  \includegraphics[width=0.4\textwidth]{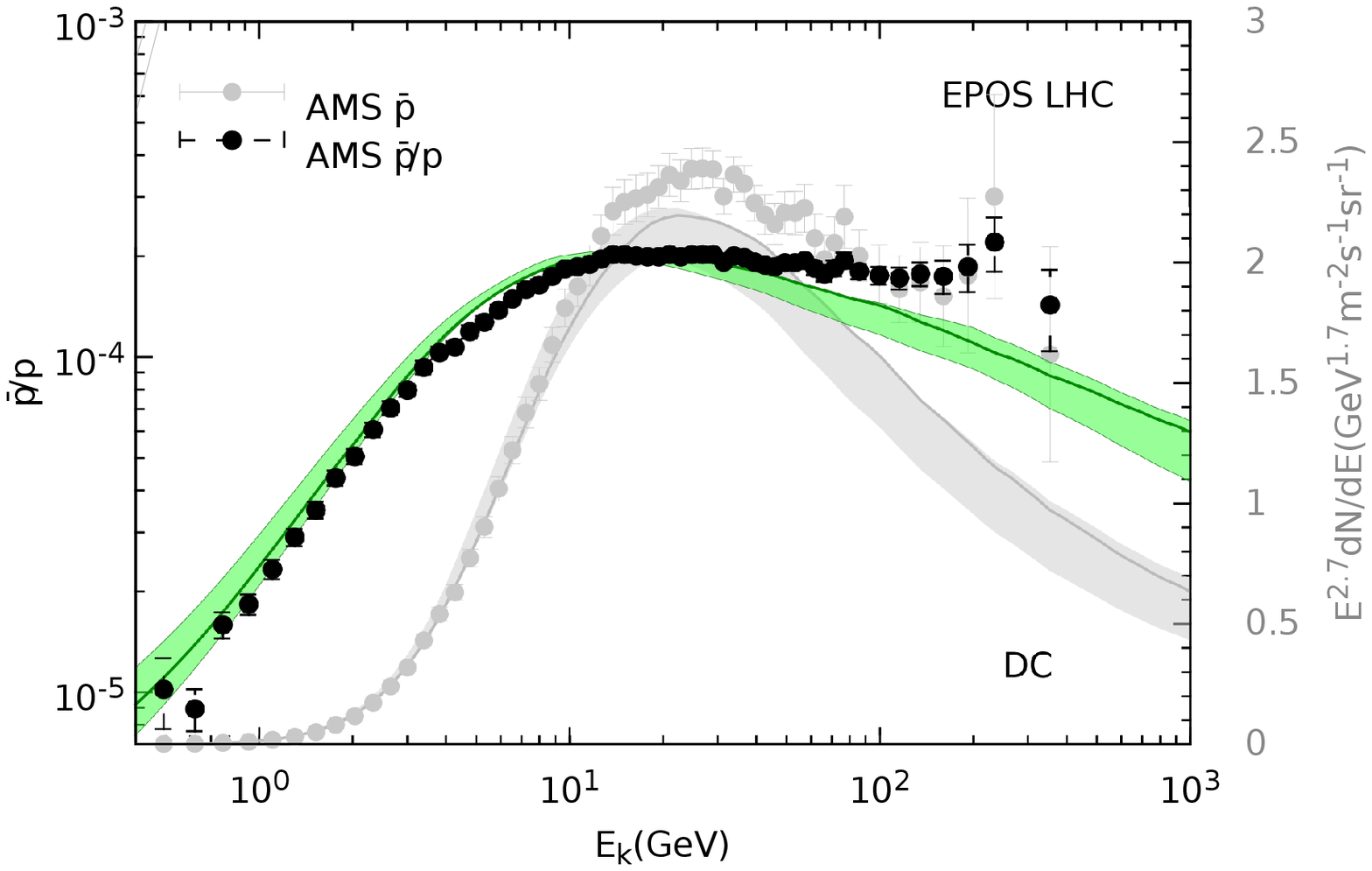}\\
  \caption{The same as \ref{fig:background_DR_cpbar} but for the DC2 model.}
  \label{fig:background_DC2_cpbar}
\end{figure}

In Figs.~\ref{fig:background_DR_cpbar}, \ref{fig:background_DR2_cpbar}, and
\ref{fig:background_DC2_cpbar}, we show the $\bar{p}/p$ ratio and $\bar{p}$ flux
for four hadronic interaction models and three CR propagation models.  To get
these result, besides the scale factor $c_{\bar{p}}$ we adopt a charge-dependent
solar modulation potential to fit the AMS-02 data better.  The colored bands of
the $\bar{p}/p$ ratio and the grey bands of $\bar{p}$ flux represent the
uncertaities of the corresponding propagation models with the propagation
parameters varying within 95\% confidence ranges.  Note that the GALPROP default
hadronic interaction model underestimates antiprotons at low energies compared
with the AMS-02 data for the DR propagation model. Such an underproduction has
been realized for a long time when studying the
PAMELA~\cite{Adriani:2008zq,Adriani:2010rc} and BESS~\cite{Orito:1999re} results
\cite{Evoli:2011id,Moskalenko:2001ya,Trotta:2010mx,Hooper:2014ysa}.  In order to
solve this problem, the diffusion coefficient at low energies should be
modified~\cite{Moskalenko:2001ya}. This is a motivation for introducing a factor
$\eta$ in the DR-2 model \cite{DiBernardo:2010is}.  Here we note that this
underproduction actually depends on the hadronic interaction models too. On the
contrary, the QGSJET-II04m model would overproduce antiprotons at low energies
in the DR propagation model.

We find that the EPOS 1.99 model almost gives the best fit to the AMS-02 data in
all three propagation models. It naturally predicts a flat $\bar{p}/p$ ratio at
the high energy end. On the other hand, the $\bar{p}$ flux predicted by the
QGSJET-II 04m model is not quite consistent with the AMS-02 data.  The EPOS LHC
model, which nearly best fit the accelerator data, seems to underestimate the
$\bar{p}/p$ ratio at the high energy end with obvious discrepancies compared to
last several data points.

In Table \ref{tab:background} we show the parameters that can give best fit of
the $\bar{p}/p$ ratio to the AMS-02 data for the different propagation and
hadronic interaction models.  The propagation parameters are within the $95\%~
C.L.$ range fitting to B/C.  Here $\phi_p$ is the solar modulation potential for
protons by fitting to the AMS-02 proton spectrum.  The solar modulation
potential value for antiproton $\phi_{\bar{p}}$ is only slightly different from
$\phi_p$.  We notice that the rescale factor $c_{\bar{p}}$ is near 1 within
20\%.  It can be seen that the EPOS 1.99 and the GALPROP default models can well
fit the AMS-02 data with small $\chi^2$ values. EPOS LHC also provides an
acceptable fit to the data, while QGSJET II-04m tends to give a large $\chi^2$.

\begin{table}
  \centering
  \input{bkg_table.tex}
  \caption{The values of parameters for the best fit to data in different scenarios.}
  \label{tab:background}
\end{table}

\section{Implications for dark matter annihilation}

As shown in the previous section, the theoretical prediction of the antiproton
flux has quite large uncertainties.  In some cases the prediction fit the AMS-02
data very well, which can give a strong constraint on the dark matter
annihilation.  However, in some cases, there are discrepancies between the
prediction and the data, which may indicate a contribution from dark matter.  We
discuss the implications on DM annihilation according to the AMS-02 antiproton
result in this section.

\begin{figure}[!htp]
  \centering
  \includegraphics[width=0.4\textwidth]{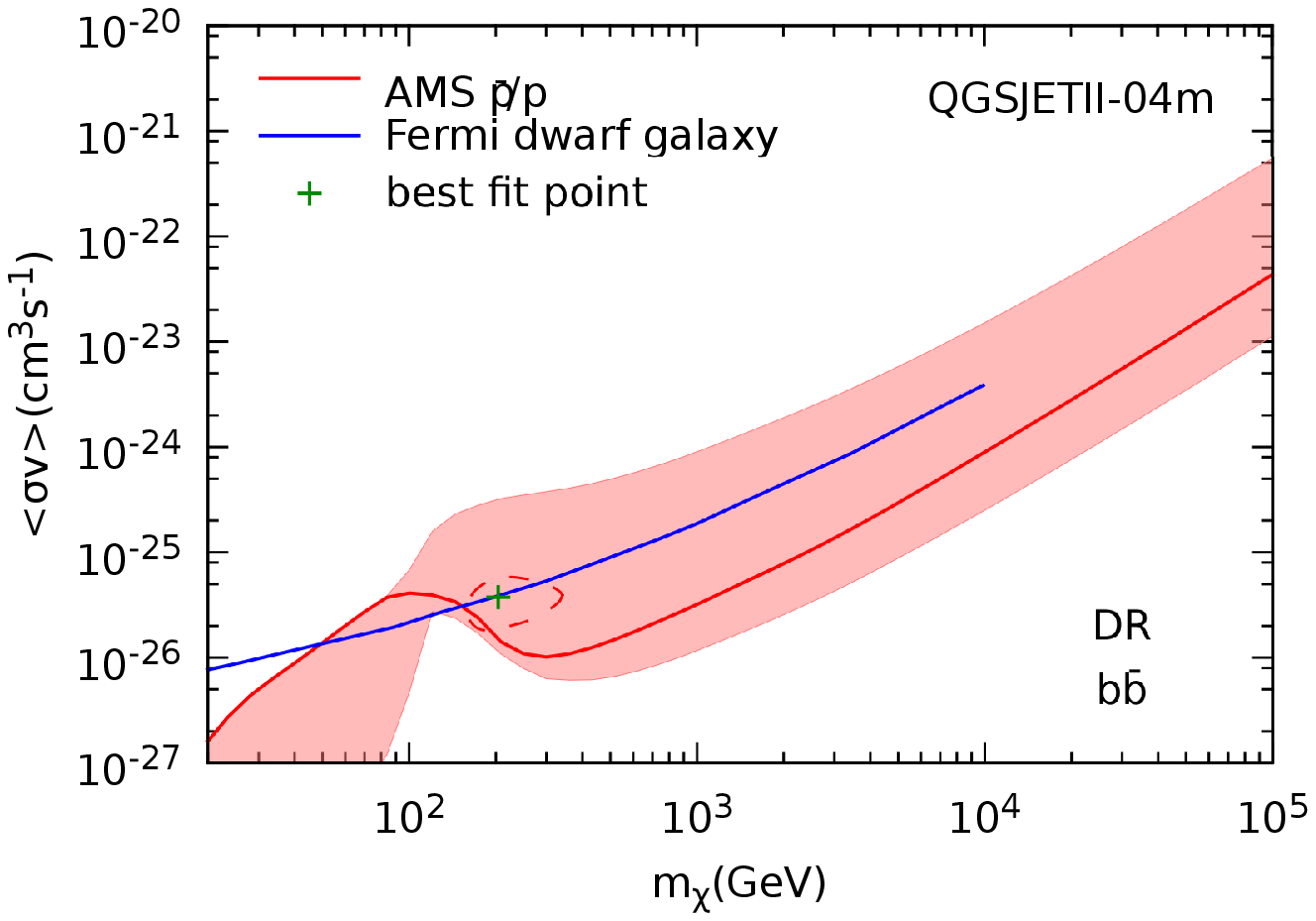}
  \includegraphics[width=0.4\textwidth]{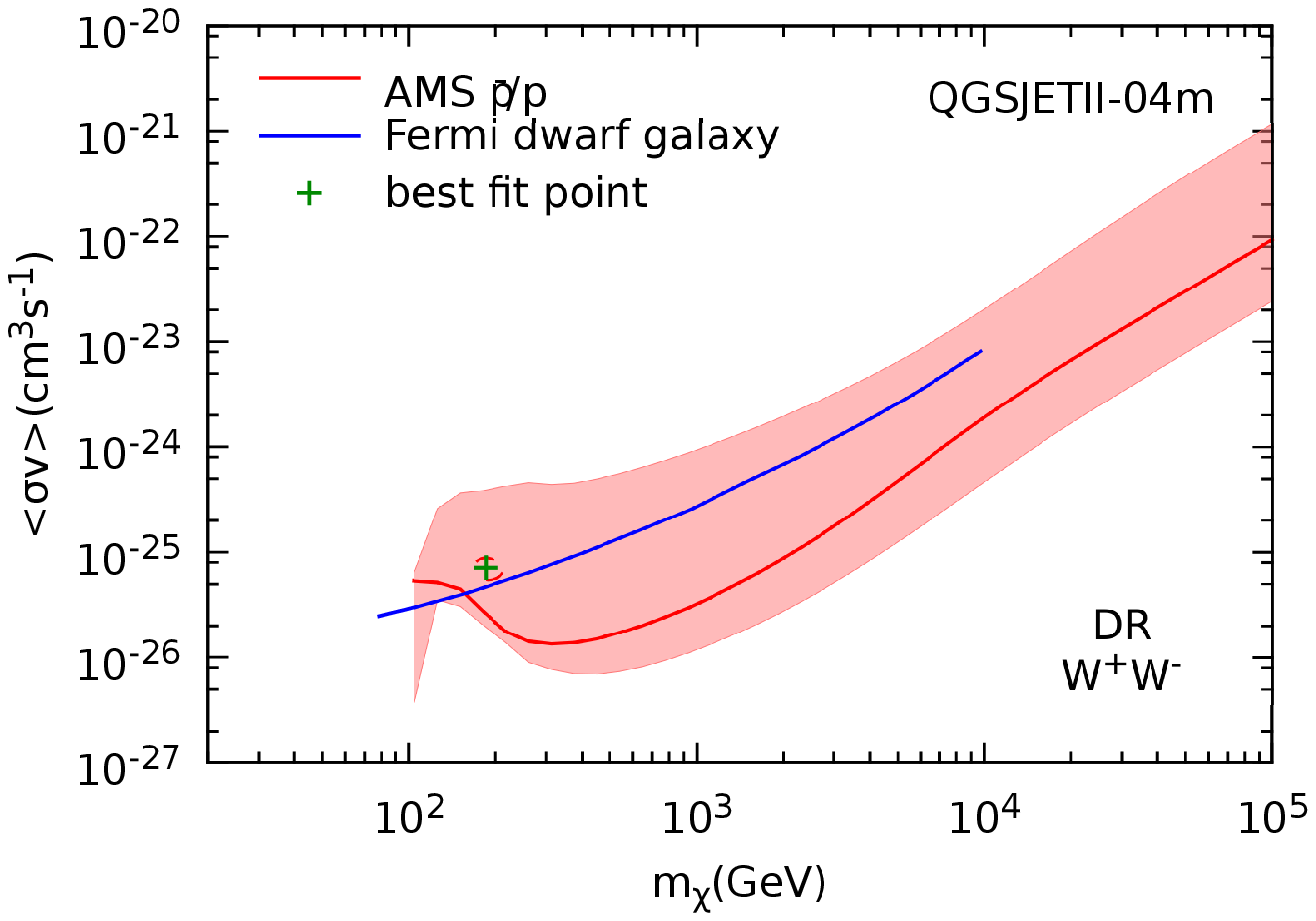}
  \caption{The $95\%$ upper limits on the dark matter annihilation rate
    $\langle\sigma v\rangle$ derived from the AMS-02 $\bar{p}/p$ ratio for
    QGSJET II-04m with the DR propagation model. The left panel is for the
    $\bar{b}b$ channel, and the right panel is for the $W^+W^-$ channel. The red
    bands indicate the propagation uncertainty in 95\% C.L., while the red line
    is corresponding to the case that fit to the $\bar{p}/p$ best. For
    comparison, we also show the upper limits from the Fermi-LAT observation of
    dwarf galaxies \cite{Ackermann:2015zua} as the blue lines. The contour
    denotes a DM signal in the $68\%$ confidence region favored by the AMS-02
    $\bar{p}/p$ ratio data in the $m_{\chi}-\langle \sigma v\rangle$ plane.}
  \label{fig:qgsjet04_constraint}
\end{figure}

\begin{figure}[!htp]
  \centering
  \includegraphics[width=0.4\textwidth]{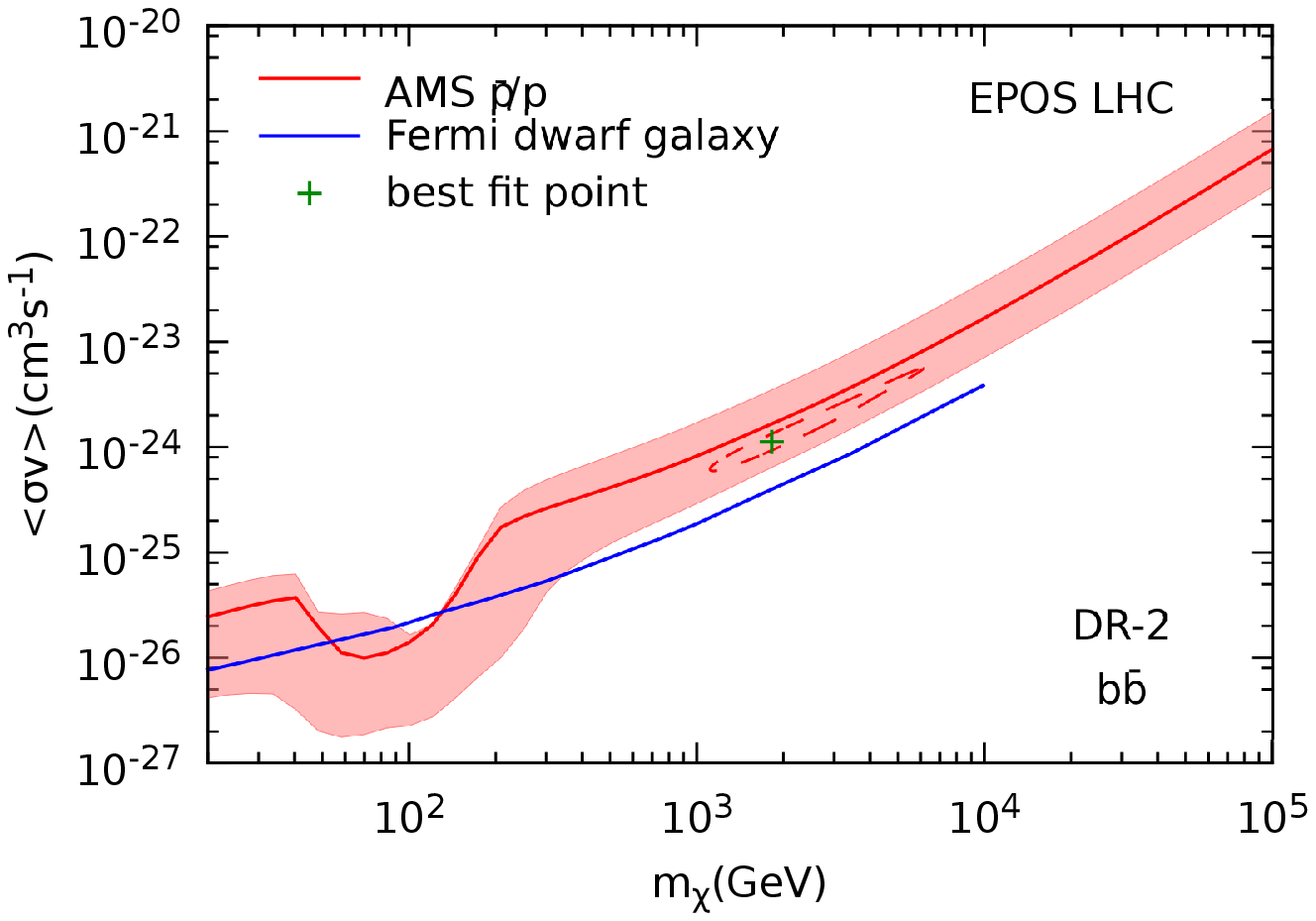}
  \includegraphics[width=0.4\textwidth]{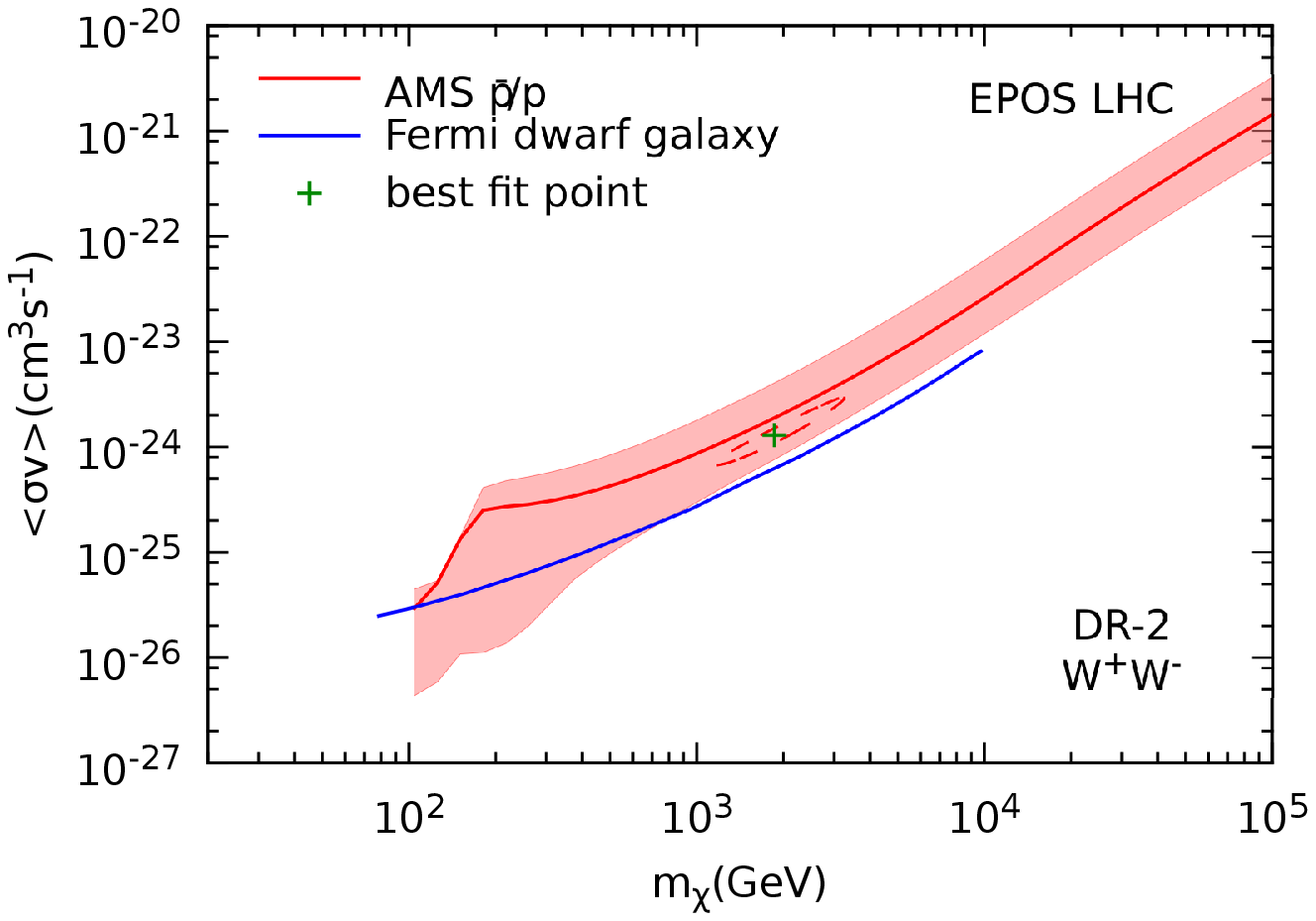}\\
  \includegraphics[width=0.4\textwidth]{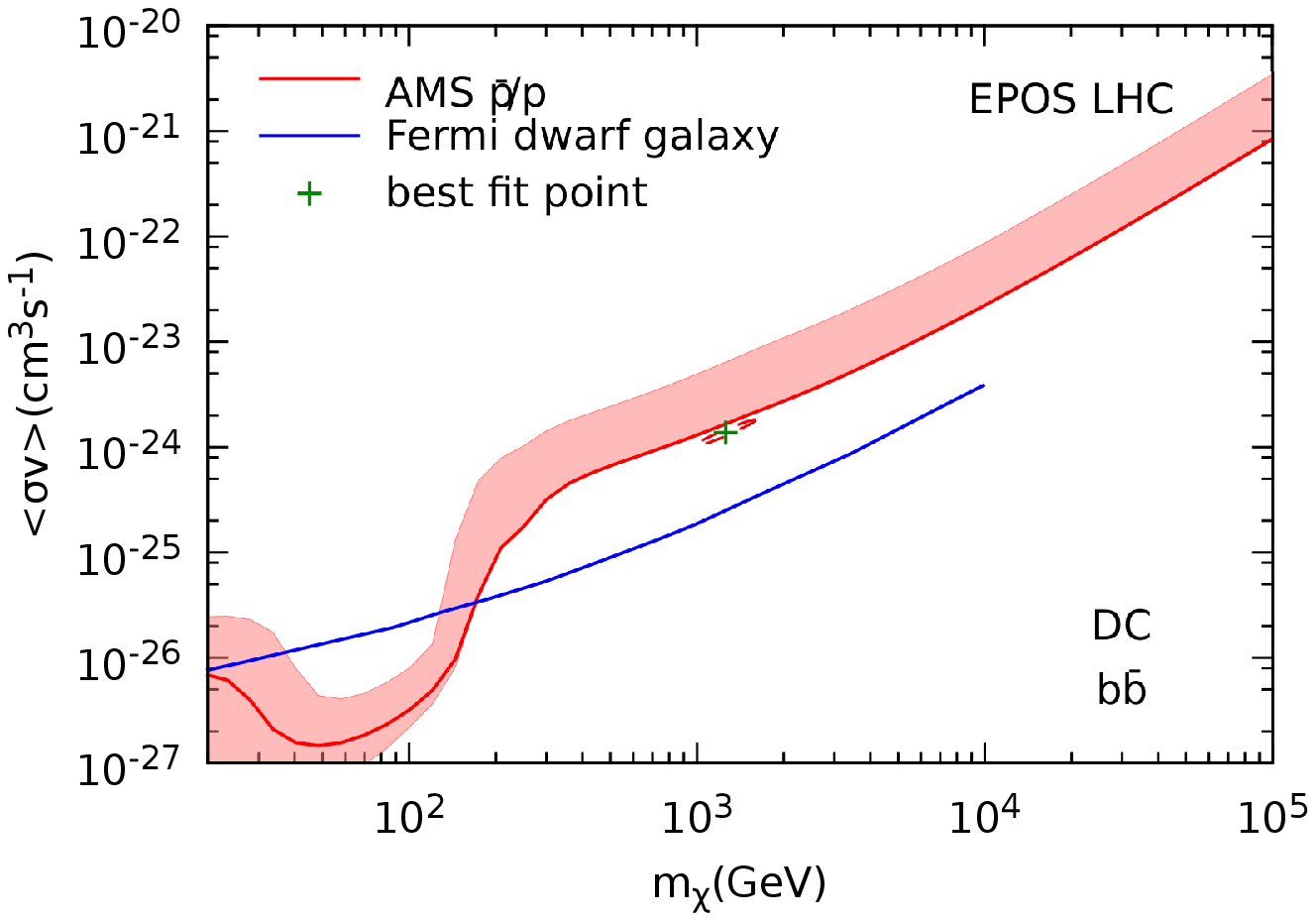}
  \includegraphics[width=0.4\textwidth]{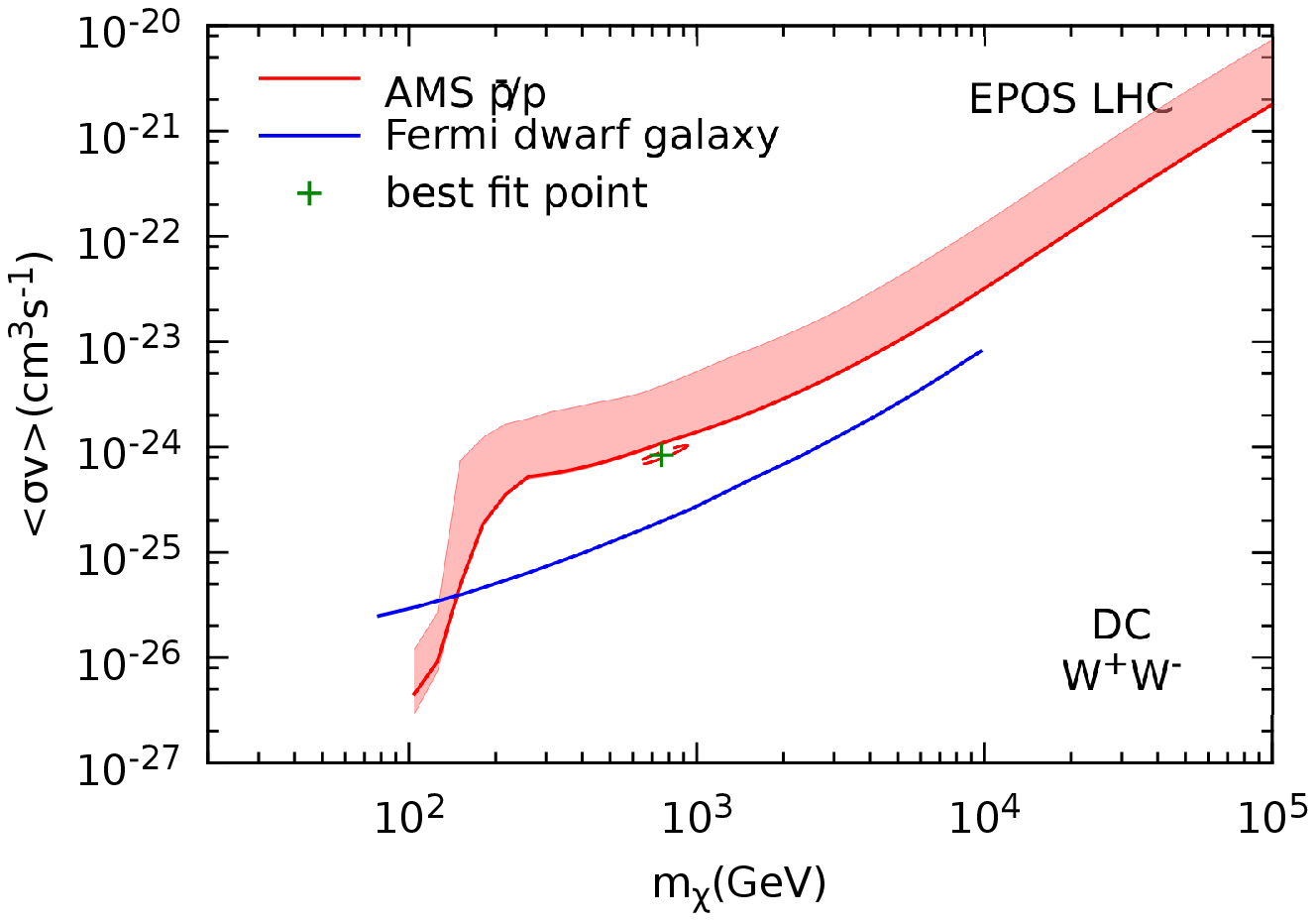}
  \caption{The same as Fig.~\ref{fig:qgsjet04_constraint}, but for EPOS LHC interaction model with the DR-2 and DC propagation models.}
  \label{fig:epos_lhc_constraint}
\end{figure}

\begin{figure}[!htp]
  \centering
  \includegraphics[width=0.4\textwidth]{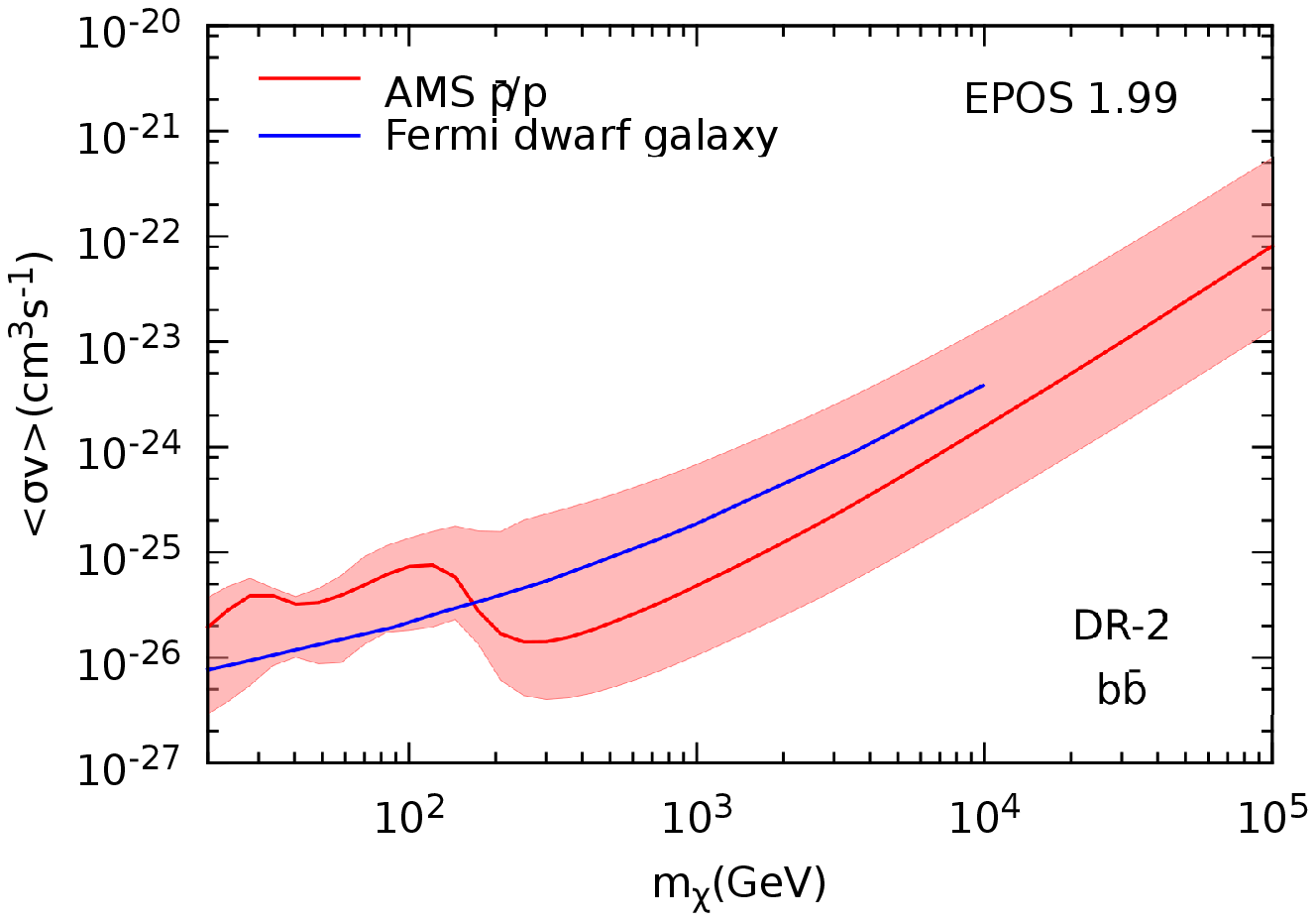}
  \includegraphics[width=0.4\textwidth]{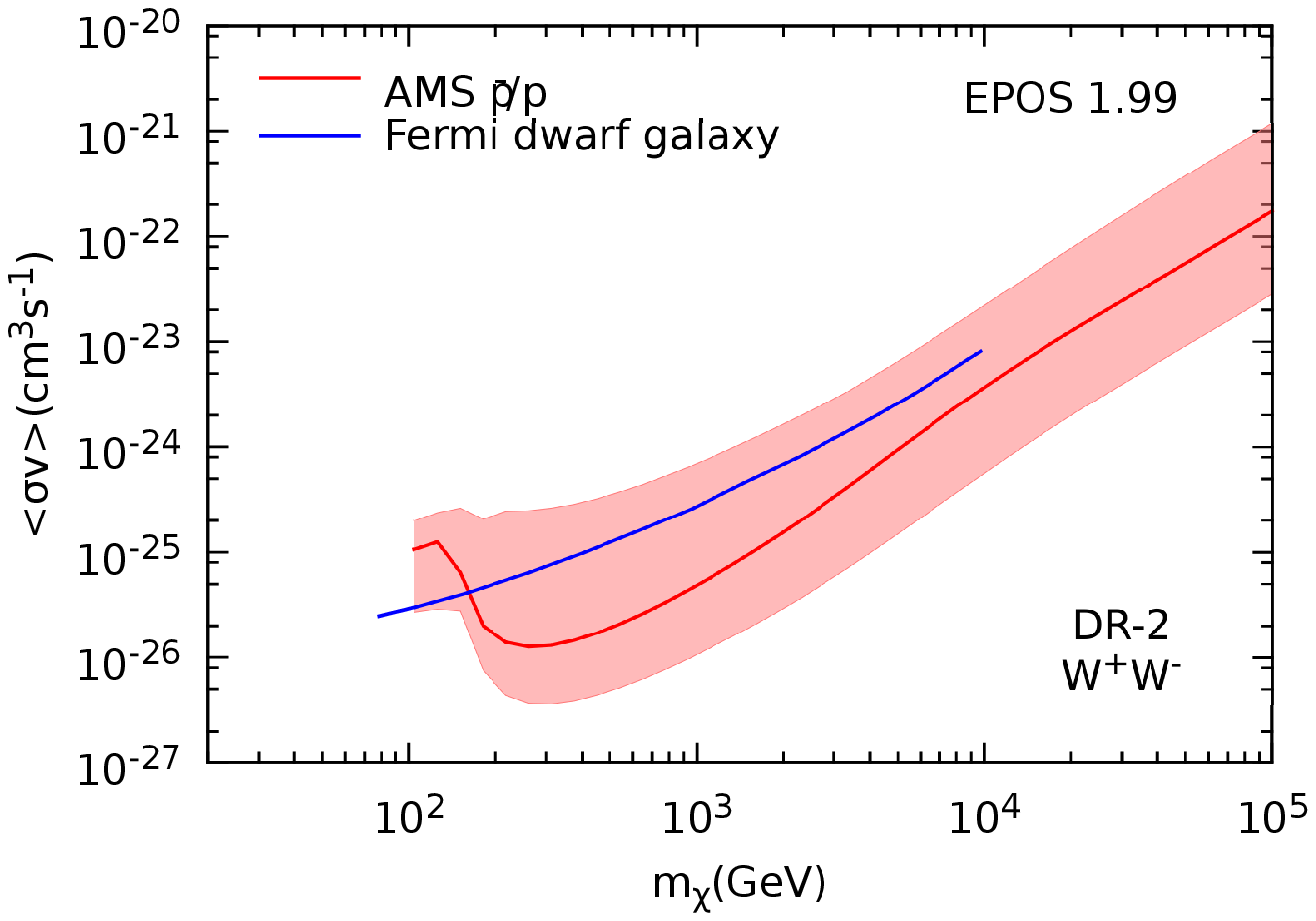}\\
  \includegraphics[width=0.4\textwidth]{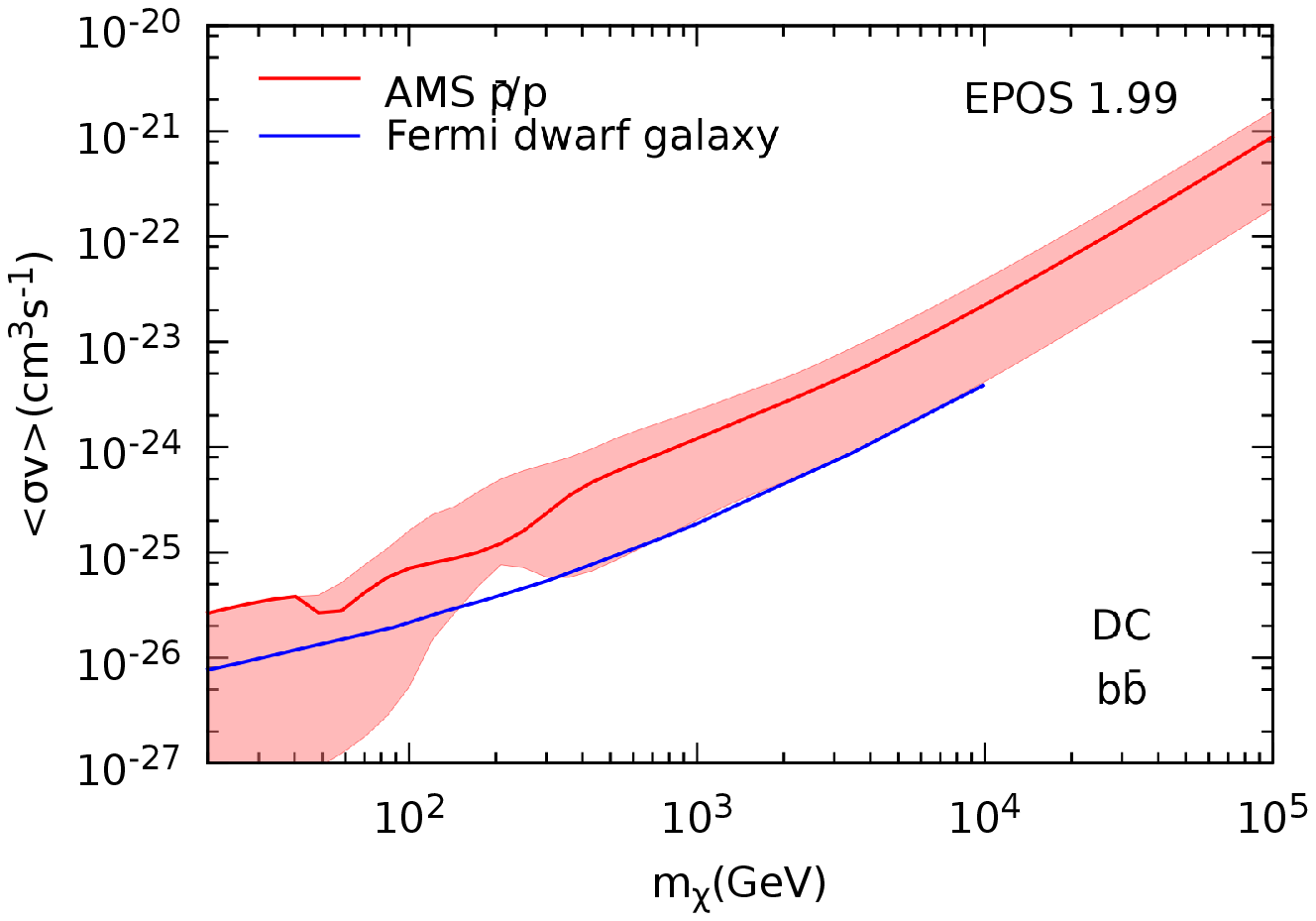}
  \includegraphics[width=0.4\textwidth]{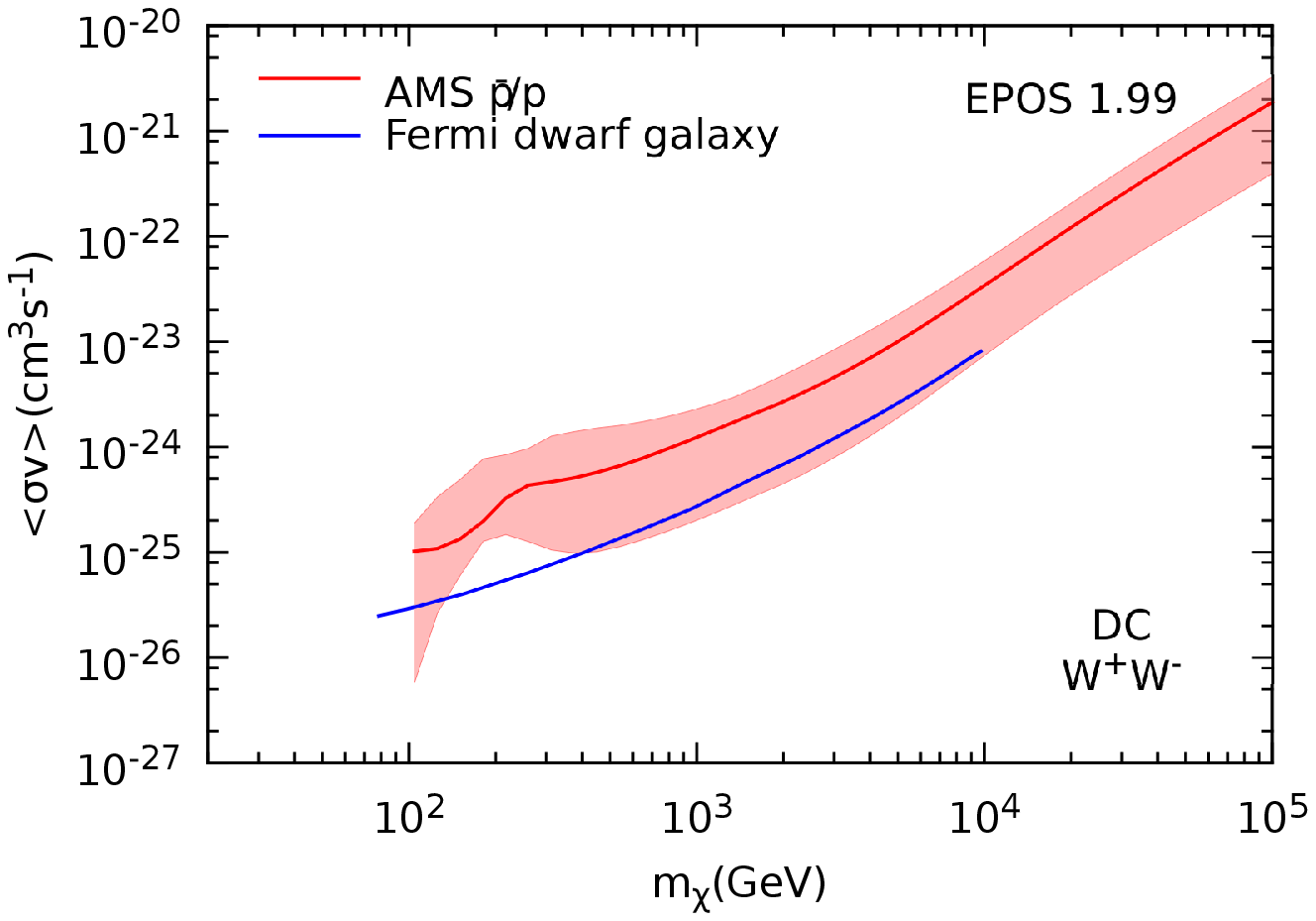}
  \caption{The same as Fig.~\ref{fig:qgsjet04_constraint}, but for EPOS 1.99 interaction model with the DR-2 and DC propagation models.}
  \label{fig:epos_199_constraint}
\end{figure}

We use the results of PPPC 4 DM ID \cite{Cirelli:2010xx} including the
electroweak corrections \cite{Ciafaloni:2010ti} to calculate DM annihilation.
The DM density distribution is assumed to be the Navarro-Frenk-White (NFW)
profile~\cite{Navarro:1996gj} with the local density of 0.3~GeV~cm$^{-3}$.  In
Figs.~\ref{fig:qgsjet04_constraint}, \ref{fig:epos_lhc_constraint}, and
\ref{fig:epos_199_constraint}, we set upper bounds on the DM annihilation cross
sections in the $b\bar{b}$ and $W^+W^-$ channels at 95\% C.L. by using the
AMS-02 $\bar{p}/p$ result. Note that if the discrepancies between the predicted
$\bar{p}/p$ ratio and the AMS-02 data at the high energy end are taken
seriously, there may be room for DM signals at high energies.  Therefore, we
also attempt to introduce a DM signal to improve the fitting quality, and show
corresponding regions of the parameter space in
Figs.~\ref{fig:qgsjet04_constraint}, \ref{fig:epos_lhc_constraint}, and
\ref{fig:epos_199_constraint}.

Fig.~\ref{fig:qgsjet04_constraint} shows the upper limits on the DM annihilation
cross sections for the $b\bar{b}$ and $W^+W^-$ final states for the QGSJET
II-04m model.  Here we only give the constraint in the DR propagation model,
since in the other two propagation models the secondary $\bar{p}/p$ predictions
can not fit the data to derive a reasonable bound. The red bands represent the
uncertainties from the propagation model. For comparison, the limits from the
Fermi-LAT gamma-ray observation of dwarf galaxies are also shown in the plots.
We can see that the constraints from the AMS-02 $\bar{p}/p$ result are more
stringent than the Fermi-LAT constraints.  We also find a small parameter region
located around 200 GeV, where a DM annihilation signal would improve the fitting
quality.

In Fig.~\ref{fig:epos_lhc_constraint}  we set constraints for the EPOS-LHC
model. The constraints are worse than the Fermi dwarf galaxy constraints, since
the EPOS LHC model generally predicts a slightly lower $\bar{p}/p$ ratio than
the data at the high energy end. Recalling that the EPOS LHC model fits the
accelerator data well, such discrepancies should be taken seriously.  In order
to improve the fit, a DM signal with a mass of $\sim\TeV$ can be introduced.
However, the corresponding DM annihilation signal seems to be excluded by Fermi.
Here we mention that such a contradiction may be evaded, if a velocity-dependent
DM annihilation cross section is introduced \cite{Zhao:2016xie}.

We give the constraints for the EPOS 1.99 model in
Fig.~\ref{fig:epos_199_constraint}. As the EPOS 1.99 model fits the AMS-02
antiproton data best, it sets very stringent constraints on the DM annihilation
cross section.

\section{Conclusions}

In this work we recalculate the astrophysical CR $\bar{p}/p$ ratio and $\bar{p}$
flux, and compare them with the latest AMS-02 results.  We investigate the
uncertainties from the propagation models and the hadronic interaction models.

We consider the DR, DR-2, and DC propagation models, where the prorogation
parameters are determined by fitting the B/C and
$^{10}\mathrm{Be}/^9\mathrm{Be}$ data with a MCMC algorithm.  All these models
can provide a very good fit to the B/C data.

We note that the uncertainties from propagation and hadronic interaction may
interplay with each other. For example, it has been recognized that the DR
propagation model may underestimate the $\bar{p}$ flux at low energies by
GALPROP \cite{Evoli:2011id,Moskalenko:2001ya,Trotta:2010mx,Hooper:2014ysa}.
However, we find that the $\bar{p}$ flux is even overproduced at low energies if
the QGSJET II-04 model is adopted.  For the DR-2 and DC propagation models, the
predictions given by QGSJET II-04m seem not consistent with the AMS-02 $\bar{p}$
data, while other three interaction models can fit the data quite well.

We further derive the upper bound on the DM annihilation cross section from the
AMS-02 $\bar{p}/p$ data. It is found that in most cases the bound can be
stronger than the constraint from Fermi observation of the dwarf galaxies.

Among these interaction models, the EPOS LHC model can fit the NA49 and LHC data
very well. However, this model predicts a slightly lower $\bar{p}/p$ ratio than
the AMS-02 data at the high energy end. If such a discrepancy is taken
seriously, a $\sim\TeV$ DM annihilation signal is necessary to fit the AMS-02
data.

\begin{acknowledgments}
  We thank Qiang Yuan, Sergey Ostapchenko for helpful discussions.  We
  appreciate Tanguy Peirog, Colin Baus, and Ralf Ulrich for the CRMC interface
  to access the hadronic models.  This work is supported by the National Natural
  Science Foundation of China under Grants No. 11475189, 11475191, by the 973
  Program of China under Grant No. 2013CB837000, and by the National Key Program
  for Research and Development (No. 2016YFA0400200).
\end{acknowledgments}

\bibliography{antiproton}
\end{document}

%% file: bkg_table.tex
\begin{tabular}{cccccccccccccccc}
\hline
           & \multicolumn{3}{c}{Tan \& Ng + BP01} &            &     \multicolumn{3}{c}{EPOS LHC}     &            &    \multicolumn{3}{c}{EPOS 1.99}     &            &   \multicolumn{3}{c}{QGSJETII-04m}  \\\cline{2-4}\cline{6-8}\cline{10-12}\cline{14-16}
           &     DR     &    DR2     &     DC     &            &     DR     &    DR2     &     DC     &            &     DR     &    DR2     &     DC     &            &     DR     &    DR2     &     DC    \\\hline
$c_{\bar p}$ &    1.15    &    1.05    &    0.99    &            &   0.899    &   0.959    &    0.76    &            &   0.923    &   0.853    &   0.836    &            &    1.22    &    1.2     &    1.04   \\
$\phi_{\bar p}$ &   0.429    &    0.69    &   0.773    &            &   0.468    &   0.928    &   0.809    &            &   0.462    &   0.592    &   0.823    &            &    0.58    &   0.843    &   0.823   \\
 $\chi^2$  & $77.9/57$  & $38.9/57$  & $51.6/57$  &            &  $131/57$  & $78.1/57$  &  $300/57$  &            &  $119/57$  & $28.9/57$  & $57.3/57$  &            &  $119/57$  &  $298/57$  & $1419/57$ \\\hline
$\phi_{p}$ &   0.515    &   0.745    &   0.644    &            &   0.562    &   0.773    &   0.674    &            &   0.554    &    0.71    &   0.686    &            &   0.483    &   0.703    &   0.686   \\\hline
  $D_0$    &    7.76    &    2.87    &    1.4     &            &    6.68    &    2.79    &    2.39    &            &    6.05    &    2.91    &    1.8     &            &    7.19    &    4.34    &    1.8    \\
$\delta$\footnote{The $\delta$ would be 0 for $R<R_0$ in the DC case} &   0.337    &   0.433    & $0/0.453$  &            &   0.338    &    0.42    & $0/0.478$  &            &   0.335    &   0.418    & $0/0.523$  &            &   0.324    &    0.39    & $0/0.523$ \\
  $R_0$    &    ---     &    ---     &    3.87    &            &    ---     &    ---     &    3.62    &            &    ---     &    ---     &    5.78    &            &    ---     &    ---     &    5.78   \\
    L      &    5.46    &    2.8     &    1.69    &            &    4.92    &    2.78    &    3.13    &            &    4.38    &    2.86    &    2.17    &            &    4.79    &    3.92    &    2.17   \\
  $v_A$    &    41.3    &    22.2    &    ---     &            &    34.4    &    18.9    &    ---     &            &    33.9    &    19.5    &    ---     &            &    42.5    &    24.4    &    ---    \\
$dV_c/dz$  &    ---     &    ---     &   0.326    &            &    ---     &    ---     &    2.97    &            &    ---     &    ---     &    2.68    &            &    ---     &    ---     &    2.68   \\
\hline
\end{tabular}